\documentclass[prd,preprint,tightenlines,floatfix,showpacs,preprintnumbers,nofootinbib,eqsecnum,superscriptaddress]{revtex4}

\usepackage[dvips,final]{graphicx}
 \usepackage{amssymb}
  \usepackage{amsmath}
   \usepackage{amsfonts}
    \usepackage{epsfig}
     \usepackage{bm}
      \usepackage{multirow}
       \usepackage{tabularx}

     \newcommand{\bq}{\mbox{\boldmath $q$}}
     \newcommand{\bk}{\mbox{\boldmath $k$}}

\newcommand{\p}{\partial}
\newcommand{\lumi}{\mbox{$\mathcal{L}$}}
\newcommand{\twosidep}[1]{\stackrel{\leftrightarrow}{\p^{#1}}}

\begin{document}


\title{QCD diffractive mechanism of exclusive $W^+W^-$ pair production
\\ at high energies}

\author{Piotr Lebiedowicz}
\email{Piotr.Lebiedowicz@ifj.edu.pl} \affiliation{Institute of
Nuclear Physics PAN, PL-31-342 Cracow, Poland}

\author{Roman Pasechnik}
\email{Roman.Pasechnik@thep.lu.se}
\affiliation{Department of Astronomy and Theoretical Physics,
Lund University, SE-223 62 Lund, Sweden}

\author{Antoni Szczurek}
\email{Antoni.Szczurek@ifj.edu.pl} \affiliation{University of
Rzesz\'ow, PL-35-959 Rzesz\'ow, Poland} \affiliation{Institute of
Nuclear Physics PAN, PL-31-342 Cracow, Poland}

\begin{abstract}
We discuss new diffractive mechanism of central exclusive production
of $W^+W^-$ pairs in proton-proton collisions at the LHC. We include
diagrams with intermediate virtual Higgs boson as well as quark box
diagrams. Several observables related to this process are
calculated. Predictions for the total cross section and differential
distributions in $W$-boson rapidity and transverse momentum as well
as $WW$ invariant mass are presented. We also show results for
different polarization states of the final $W^{\pm}$ bosons. 
We compare the contribution of the $\gamma \gamma \to W^+ W^-$
mechanism considered in the literature with the contribution of the
diffractive mechanism through the $gg \to W^+ W^-$ subprocess for
the different observables.
The phase space integrated diffractive contribution when separated
is only a small fraction of fb compared to 115.4 fb 
of the $\gamma\gamma$-contribution without absorption.
The latter contribution dominates at small four-momentum
transfers squared in the proton lines and in a broad range of $W^+
W^-$ invariant masses. This offers a possibility of efficient
searches for anomalous triple-boson ($\gamma W W$) and quartic-boson ($\gamma
\gamma WW$) couplings and testing models beyond the Standard Model.
We discuss shortly also the $p p \to p p \gamma \gamma$ process,
where the box contribution is very similar to that for $W^+ W^-$ 
and compare our results with recent CDF data. 
Nice agreement has been achieved without additional free parameters.
  
\end{abstract}



\maketitle

\section{Introduction}

The central exclusive production (CEP) process $pp\to p + X + p$,
where $X$ stands for a centrally produced system separated from the
two very forward protons by large rapidity gaps, has been proposed
in Refs.~\cite{SNS90,Bialas:1991wj} as an alternative way of searching for
the neutral Higgs boson (see Ref.~\cite{Albrow:2010yb} for a review).
If momenta of the outgoing protons are measured by forward
proton detectors placed at 220 m and 420 m
from the ATLAS/CMS interaction point \cite{FP420},
the mass of the $X$ system may be reconstructed~\cite{Albrow:2000na}
with very precise resolution.

The exclusive reaction $pp\to pHp$ has been
intensively studied by the Durham group \cite{Durham} in the last
decade. This study was motivated by the clean environment and
largely reduced background due to a suppression of $b\bar b$ production
as a consequence of the spin-parity conservation in the forward limit.
However, very recent precise calculations of Refs.~\cite{MPS_bbbar}
have shown that the situation with Higgs CEP background in the $b\bar b$
channel is more complicated and the signal is to a large extent shadowed by the
exclusive non-reducible continuum $b{\bar b}$ production.
In addition, reducible backgrounds from a misidentification of gluonic
jets as $b$-quark jets can be very difficult to separate
\cite{MPS2011_gg}. Since the total cross section for the Higgs CEP
is quite small and rather uncertain, the issue with the Higgs CEP is
still far from its final resolution, from both theoretical and
experimental point of view.

The final system $X$ in the midrapidity region is predominantly
produced in the $J_z=0$ state as dictated by the well-known $J_z=0$
selection rule \cite{Durham}. However, corrections to this rule due
to slightly off-forward protons can be important for lower (a few GeV)
mass central systems and may lead to sizeable contributions in the observable
signals, in particular, in the $\chi_c$ mesons \cite{chic,LKRS10},
$b\bar b$ \cite{MPS_bbbar} and $gg$ \cite{Cudell:2008gv,MPS2011_gg} CEP.
The emission of gluons from the "screening" gluon
could also violate the $J_z=0$ selection rule as has
recently been emphasized in Ref.~\cite{Cudell:2008gv}.

In order to reduce the theoretical uncertainties
of the CEP mechanism, coming from both the hard subprocess
(Sudakov form factor \cite{Cudell:2008gv,Dechambre:2011py},
next-to-leading order QCD corrections \cite{Khoze:2006um})
and the soft interactions
(color screening effects at extremely small gluon $x$ \cite{soft-col},
rapidity gap survival factor \cite{SF},
poorly known unintegrated gluon distribution functions (UGDFs) at small gluon
$q_\perp$ and $x$ \cite{chic,LKRS10}),
new experimental data 
on various exclusive production channels are certainly required
and expected to come soon from ongoing LHC measurements. 
In particular, as it was stressed e.g. in Ref.~\cite{Dechambre:2011py}
the measurements of the exclusive dijets production at the LHC could
largely reduce the theoretical uncertainty in the Higgs boson CEP.
Other measurements, e.g.
heavy quarkonia \cite{chic,LKRS10},
$\gamma\gamma$ \cite{LKRS10},
high-$p_\perp$ light mesons \cite{Szczurek:2006bn,HarlandLang:2011qd},
associated charged Higgs $H^+W^-$ \cite{EP2011} CEP, etc.,
are also important in this context.
Some of these results have been compared to
experimental data from the Tevatron~\cite{CDF,CDF_gamgam},
and a rough quantitative agreement between them has been achieved.

In this paper, we focus on exclusive production of $W^+W^-$ pairs 
in high-energy proton-proton collisions.
It was found recently \cite{royon,piotrzkowski} that the reaction is an 
ideal case to study experimentally $\gamma W^+ W^-$ and 
$\gamma \gamma W^+ W^-$ couplings
\footnote{Some more subtle aspects of the beyond Standard Model
anomalous couplings were discussed e.g. in \cite{MMN08}.}.

The $\gamma \gamma \to W^+ W^-$ process is interesting reaction to 
test the Standard Model and any other theory beyond the Standard Model.
The linear collider would be a good option to study the couplings of 
gauge bosons in the distant future.
For instance in Ref.\cite{NNPU} the anomalous coupling in locally 
SU(2) $\times$ U(1) invariant effective Lagrangian was studied. 
Other models also lead to anomalous gauge boson coupling. 

The photon-photon contribution for the purely exclusive production 
of $W^+ W^-$ was considered so far in the literature.
The diffractive production and decay of Higgs boson into the $W^+W^-$ pair 
was discussed in Ref.~\cite{WWKhoze}, and the corresponding cross
section turned out to be significantly smaller than that for 
the $\gamma\gamma$-contribution.
Provided this is the case, the $W^+W^-$ pair production signal would
be particularly sensitive to New Physics contributions in 
the $\gamma \gamma \to W^+ W^-$ subprocess \cite{royon,piotrzkowski}. 
Similar analysis has been considered recently
for $\gamma \gamma \to Z Z$ \cite{Gupta:2011be}. 
These previous analyses strongly motivate our present detailed study 
on a competitive diffractive contribution.
The $pp\to pW^+W^-p$ process going through the diffractive QCD
mechanism with the $gg \to W^+W^-$ subprocess naturally constitutes
a background for the exclusive electromagnetic 
$pp\to p(\gamma\gamma\to W^+W^-)p$ process.
We consider not only the mechanism with intermediate Higgs boson but
also quark box contributions never estimated in exclusive processes.
Both the Higgs and box contribution may interfere together.
We discuss here the interference effects.
Corresponding measurements will be possible to perform at the ATLAS
detector with the use of very forward proton detectors \cite{royon}. 
In order to quantify to what extent the QCD mechanism competes with the
``signal'' from the $\gamma\gamma$ fusion, we calculate both
contributions and compare them differentially
as a function of several relevant kinematical variables.

Since the box contribution of exclusive diffractive $p p \to p p W^+ W^-$
process is very similar to the $p \bar{p} \to p \bar{p} \gamma \gamma$ process
which has been measured recently \cite{CDF_gamgam}, we discuss the
latter one and compare corresponding results with the recent CDF data.

\section{Diffractive mechanism of exclusive $W^+W^-$ pair production}

\begin{figure}[h!]
\centerline{\epsfig{file=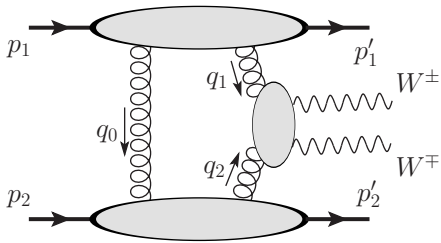,width=6.0cm}} \caption{Generic
diagram for the central exclusive $WW$ pair production in $pp$
collisions. Momenta of incident particles are shown explicitly.}
\label{fig:WWCEP}
\end{figure}

A schematic diagram for central exclusive production of
$W^{\pm}W^{\mp}$ pairs in proton-proton scattering $pp\to
pW^{\pm}W^{\mp}p$ is shown in Fig.~\ref{fig:WWCEP}. Similar
mechanisms have been considered in inclusive production of $W^+ W^-$
pairs (see e.g. Refs.~\cite{inclusive,gamgam_WW,DDS95}). In what
follows, we use the standard theoretical description of CEP
processes developed by Khoze, Martin and Ryskin for the exclusive
production of Higgs boson \cite{Durham}.

The momenta of intermediate gluons are given by Sudakov
decompositions in terms of the incoming proton four-momenta
$p_{1,2}$
\begin{eqnarray}\nonumber
&&q_1=x_1p_1+q_{1\perp},\quad q_2=x_2p_2+q_{2\perp},\quad 0<x_{1,2}<1,\\
&&q_0=x'p_1-x'p_2+q_{0\perp}\simeq q_{0\perp},\quad x'\ll x_{1,2},
\label{moms}
\end{eqnarray}
where $x_{1,2},x'$ are the longitudinal momentum fractions for
active (fusing) and color screening gluons, respectively.

In the forward proton scattering limit, we have
\begin{eqnarray}\nonumber
&&t_{1,2}=(p_{1,2}-p'_{1,2})^2\simeq{p'}^2_{1,2\perp}\to0 \,,\\
&&q_{\perp} \equiv q_{0\perp} \simeq -q_{1\perp} = q_{2\perp} \,.
\label{forward}
\end{eqnarray}
The QCD factorisation of the process at the hard scale $\mu_F$ is
provided by the large invariant mass of the $WW$ pair $M_{WW}$, i.e.
\begin{eqnarray}\label{sx1x2}
\mu_F^2\equiv s\,x_1x_2\simeq M_{WW}^2\,.
\end{eqnarray}
It is convenient to introduce the Sudakov expansion for $W^{\pm}$
boson momenta
\begin{eqnarray}
k_+=x_1^+ p_1+x_2^+ p_2+k_{+\perp},\quad
k_-=x_1^- p_1+x_2^- p_2+k_{-\perp}
\end{eqnarray}
leading to
\begin{eqnarray}\label{xqq}
x_{1,2}=x_{1,2}^+ + x_{1,2}^-,\quad
x_{1,2}^+=\frac{m_{+\perp}}{\sqrt{s}}e^{\pm y_+},\quad
x_{1,2}^-=\frac{m_{-\perp}}{\sqrt{s}}e^{\pm y_-},\quad
m_{\pm\perp}^2=m_W^2+|{\bk}_{\pm\perp}|^2\,,
\end{eqnarray}
in terms of $W^{\pm}$ rapidities $y_{\pm}$ and transverse masses
$m_{\pm\perp}$. For simplicity, in actual calculations we work in
the forward limit given by Eq.~(\ref{forward}), which implies that
${\bk}_{+\perp}=-{\bk}_{-\perp}$.

In actual calculations below, $W^{\pm}$ bosons are assumed to be
on-mass-shell, whereas particular contributions to the observables
can then be estimated in the narrow-width approximation. For
example, in the leptonic channel we have the following observable
cross section
\begin{eqnarray}
\sigma_{l^+\nu l^-\nu}\simeq \sigma_{WW}\times \mbox{BR}(W^+\to
l^+\nu)\,\mbox{BR}(W^-\to l^-\nu) \, ,
\end{eqnarray}
where $\mbox{BR}(W^+ \to l^+ \nu) = (10.80 \pm 0.09) \times 10^{-2}$
\cite{PDG} for a given lepton flavor. Both electrons and muons can
be used in practice \cite{royon}.

We write the amplitude of the diffractive process, which at high energy
is dominated by its imaginary part, as
\begin{eqnarray} \label{ampl}
{\cal M}_{\lambda_+\lambda_-}(s,t_1,t_2) &\simeq&is\frac{\pi^2}{2}
\int d^2 {\bq}_{0\perp} V_{\lambda_+\lambda_-}(q_1,q_2,k_{+},k_{-})
\frac{f_g(q_0,q_1;t_1)f_g(q_0,q_2;t_2)}
{{\bq}_{0\perp}^2\,{\bq}_{1\perp}^2\,{\bq}_{2\perp}^2}\,,
\end{eqnarray}
where $\lambda_{\pm}=\pm 1,\,0$ are the polarisation states
of the produced $W^{\pm}$ bosons, respectively, $f_g(r_1,r_2;t)$ is
the off-diagonal unintegrated gluon distribution function (UGDF),
which depends on the longitudinal
and transverse components of both gluons momenta.
The gauge-invariant $gg\to W_{\lambda_+}^+ W_{\lambda_-}^-$
hard subprocess amplitude
$V_{\lambda_+\lambda_-}(q_1,q_2,k_{+},k_{-})$
is given by the light cone projection
\begin{eqnarray}\label{GIproj}
V_{\lambda_+\lambda_-}=
n^+_{\mu}n^-_{\nu}V_{\lambda_+\lambda_- , \mu\nu}=
\frac{4}{s}
\frac{q^{\mu}_{1\perp}}{x_1}
\frac{q^{\nu}_{2\perp}}{x_2}
V_{\lambda_+\lambda_-,\mu\nu},\quad
q_1^{\mu}V_{\lambda_+\lambda_-,\mu\nu}=
q_2^{\nu}V_{\lambda_+\lambda_-,\mu\nu}=0\,,
\end{eqnarray}
where $n_{\mu}^{\pm} = p_{1,2}^{\mu}/E_{p,cms}$ and the
center-of-mass proton energy $E_{p,cms} = \sqrt{s}/2$. We adopt the
definition of gluon transverse polarisation vectors proportional to
the transverse gluon momenta $q_{1,2 \perp}$, i.e. $\epsilon_{1,2} \sim q_{1,2 \perp} /
x_{1,2}$. The helicity matrix element in the previous expression reads
%
\begin{eqnarray}
V_{\lambda_+\lambda_-}^{\mu\nu}(q_1,q_2,k_{+},k_{-})=
\epsilon^{*,\rho}(k_+,\lambda_+)
\epsilon^{*,\sigma}(k_-,\lambda_-)V_{\rho\sigma}^{\mu\nu}\,,
\label{Vepsilon}
\end{eqnarray}
in terms of the Lorentz and gauge invariant $2\to2$ amplitude
$V_{\rho\sigma}^{\mu\nu}$ and $W$ boson polarisation vectors
$\epsilon(k,\lambda)$. Below we will analyze the exclusive
production with polarized $W^+$ or $W^-$. In Eq.~(\ref{Vepsilon})
$\epsilon_{\mu}(k_+,\lambda_+)$ and
$\epsilon_{\nu}(k_-,\lambda_-)$ can be defined easily in the
proton-proton center-of-mass frame 
as
\begin{eqnarray}
\epsilon(k,0) &=& \frac{E_{W}}{m_{W}}
\left(\frac{k}{E_{W}},\,\cos\phi \sin\theta, \,
\sin\phi \sin\theta,\,\cos\theta\right) \,,\nonumber \\
\epsilon(k,\pm 1) &=& \frac{1}{\sqrt{2}} \left(0,\, i\sin\phi \mp
\cos\theta\cos\phi,\, -i\cos\phi \mp \cos\theta\sin\phi,\, \pm
\sin\theta\right)\,,
\label{vectors}
\end{eqnarray}
where $\phi$ is the azimuthal angle of a produced boson, and satisfy
$\epsilon^{\mu}(\lambda)\epsilon^*_{\mu}(\lambda)=-1$ and
$\epsilon^*_{\mu}(k_+,\lambda_+)k_+^{\mu}=\epsilon^*_{\nu}(k_-,\lambda_-)k_-^{\nu}=0$.
In the forward limit, provided by Eq.~(\ref{forward}), the azimuthal
angles of the $W^+$ and $W^-$ bosons are related as $\phi_{-} = \phi_{+} + \pi$.

The diffractive amplitude given by Eq.~(\ref{ampl}) is averaged over
the color indices and over the two transverse polarizations of the
incoming gluons. The relevant color factor which includes summing
over colors of quarks in the loop (triangle or box) and averaging
over fusing gluon colors (according to the definition of
unintegrated gluon distribution function) is the same as in the
previously studied Higgs CEP (for more details on derivation of the
generic $pp\to pXp$ amplitude, see e.g. Ref.~\cite{Albrow:2010yb}).
The matrix element $V_{\lambda_{+},\lambda_{-}}$ contains twice the
strong coupling constant $g_s^2 = 4 \pi \alpha_{s}$. In our
calculation here we take the running coupling constant
$\alpha_s(\mu_{hard}^2=M_{WW}^2)$ which depends on the invariant
mass of $WW$ pair as a hard renormalisation scale of the process.
The choice of the scale approximately introduces roughly a factor of
two model uncertainties when varying the hard scale $\mu_{hard}$
between $2M_{WW}$ and $M_{WW}/2$ values.

The bare amplitude above is subjected to absorption corrections that
depend on the collision energy and typical proton transverse
momenta. As in the original KMR calculations \cite{Durham}, the bare
production cross section is usually multiplied by a rapidity gap survival
factor which we take the same as for the Higgs boson and $b \bar b$
production to be $S_{g} = 0.03$ at the LHC energy (see e.g.
Ref.~\cite{MPS2011_gg}).

\subsection{The hard subprocess}

\begin{figure}[h!]
\centerline{\epsfig{file=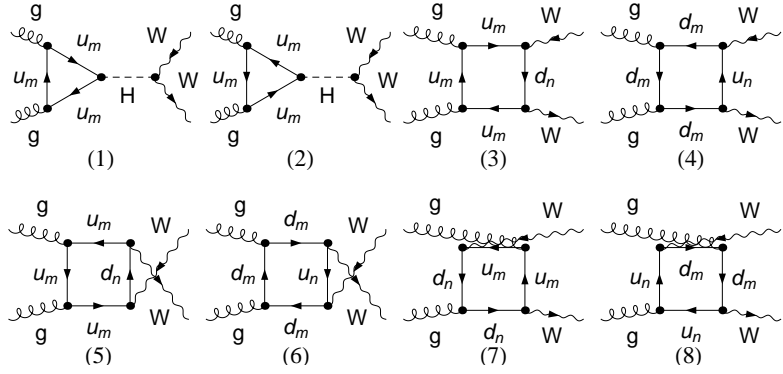,width=10.5cm}}
\caption{Representative diagrams of the hard subprocess
$gg\to W^{\pm}W^{\mp}$,
which contribute to the exclusive $WW$ pair production.}
\label{fig:WWhard}
\end{figure}

The typical contributions to the $gg\to W^+W^-$ subprocess are shown
in Fig.~\ref{fig:WWhard}. The total number of topologically
different loop diagrams amounts to two triangles, and six boxes. In
the central exclusive $W^+W^-$ production, triangle diagrams with
$\gamma$ and $Z$ bosons in the intermediate state are suppressed due
to the $J_z$ = 0 and parity selection rule for singlet gluon-gluon to 
(virtual) photon transition strictly valid in the on-shell limit of 
fusing gluons and Landau-Yang theorem for intermediate $Z$ boson.

Then the only non-zeroth contribution comes from the Higgs resonant
diagram, and in the next subsection we will discuss it in detail.
However, this can only lead to a sizeable enhancement of the cross
section close to its threshold $m_{h^0}\simeq M_{WW}\gtrsim 2 m_W$
\cite{WWKhoze}. The Standard Model Higgs bosons with such large
masses have been recently excluded by the Tevatron
\cite{CDF_exclusion} and LHC \cite{ATLAS_exclusion,CMS_exclusion}
measurements. For yet allowed values of Higgs mass 115 GeV $\lesssim
m_{h^0} \lesssim $ 130 GeV, corresponding contribution to the
$W^{+}W^{-}$ channel is far from the Higgs boson resonance and turned
out to be suppressed compared to box contributions at low invariant masses.
However, due to interference effects at rather large
invariant masses $M_{WW}$ the resonant (triangles) contribution
could become comparable to the non-resonant (boxes) one. Below, for
comparison we have calculated box and triangle (through the
$s$-channel SM Higgs boson exchange) contributions in different phase
space regions 
\footnote{Close to the $WW$-threshold instability
of $W$ bosons \cite{Wsmear} should be included.}
which could be interesting for future measurements
with forward detectors at ATLAS or CMS.

\subsubsection{Higgs contribution}

The matrix element for the $gg \to h^0 \to W^+ W^-$ transition with
intermediate $s$-channel Higgs boson exchange
(see first two diagrams in Fig.~\ref{fig:WWhard})
can be written in the narrow-width approximation as
\begin{eqnarray}\nonumber
&&V_{gg\to h^0\to
W^+W^-}(q_1,q_2,k_+,k_-)=\delta^{(4)}(q_1+q_2-k_+-k_-)\times \\
&&\qquad V_{gg \to h^0}(q_1,q_2,p_{h^0}) \, \frac{i}{M_{WW}^2 -
m_{h^0}^2 + iM_{WW}\Gamma_{\rm{tot}}^h} \, V_{h^0 \to
W^+W^-}(k_+, k_-,\lambda_+,\lambda_-),
\label{triangle}
\end{eqnarray}
where the Higgs boson momentum is $p_{h^0}=q_1+q_2$, and the
$\delta$-function reflects the momentum conservation in the process.
In order to get a correct resonant invariant mass distribution, the
standard Breit-Wigner Higgs propagator with the total Higgs decay
width $\Gamma^{h}_{\rm{tot}}$, which can be found e.g. in
Ref.~\cite{Passarino}, is used.

In Eq.~(\ref{triangle}), first the $gg\to h^0$ amplitude of the Higgs
boson production through the top-quark triangle in the $k_t$-factorisation approach
can be written as (see e.g. Ref.~\cite{incl-Higgs})
\begin{equation}
V_{gg \to h^0}\simeq
\frac{i\delta^{ab}}{v}\,\frac{\alpha_s(\mu_F^2)}{\pi}\,(\bq_{1\perp}\cdot
\bq_{2\perp})\,\frac{2}{3}\left(1+\frac{7}{120}\frac{M_{WW}^2}{m_{\rm{top}}^2}\right)\,,\qquad
v=\left(G_F\sqrt{2}\right)^{-1/2}.
\end{equation}
The second tree-level $h^0 \to W^+ W^-$ ``decay'' amplitude reads:
\begin{equation}
V_{h^0 \to W^+W^-}\simeq i m_W\,\frac{e}{\sin\theta_W}\,
\epsilon^*(k_+,\lambda_+) \epsilon^*(k_-,\lambda_-)\,,
\end{equation}
where the polarisation vectors in the direction of motion of $W^+$
and $W^-$ bosons in the proton-proton center-of-mass frame are used
in practical calculations.

Potentially interesting contribution could come from the Higgs
resonance if the Higgs mass was close to the $WW$ production
threshold. Similar resonance effects have been considered recently
in inclusive \cite{Enberg:2011ae} and exclusive associated
\cite{EP2011} charged Higgs boson production, and large
contributions beyond the Standard Model were found. However, the SM
Higgs mass $\sim$160 GeV has been recently excluded in inclusive
searches by the CDF Collaboration at Tevatron \cite{CDF_exclusion}
and by the ATLAS and CMS Collaborations at LHC
\cite{ATLAS_exclusion,CMS_exclusion}, so yet realistic SM Higgs
boson mass interval $m_{h^0}\sim 115-130$ GeV leads to a suppressed triangles'
contribution to exclusive $W^{+} W^{-}$ pair production.
In the calculation presented here we take $m_{h^{0}}$ = 120 GeV.
Since the Higgs mass is certainly much smaller than the threshold value
a precise value of the Higgs boson mass is not very important.
A contribution from an extended Higgs sector
beyond the Standard Model \cite{Enberg:2011ae} could be
interesting, but we postpone this issue for a later study.

In this work, we are primarily interested in estimation of dominant
box contributions as well as in possible box-triangle interference
effects within the Standard Model as potentially important irreducible
background for the $\gamma\gamma\to W^+W^-$ signal relevant for a
precision study of anomalous couplings. Thus, our numerical
estimates provide minimal limit for the central exclusive $WW$
production signal.

\subsubsection{Contribution of box diagrams}

The box contributions to the $gg\to W^+W^-$ parton level subprocess
amplitude (see diagrams No.~(3-8) in Fig.~\ref{fig:WWhard}) for
on-shell fusing gluons were calculated analytically by using the
Mathematica-based {\tt FormCalc} (FC) \cite{FC} package. The
complete matrix element was generated automatically by the FC tools in
terms of one-loop Passarino-Veltman two-, three- and four-point
functions and other internally-defined functions (e.g. gluon and
vector bosons polarisation vectors) and kinematical variables.

At the next step, the Fortran code for the matrix element was
generated, and then used as an external subroutine in our numerical
calculations together with other FC routines setting up the Standard
Model parameters, coupling constants and kinematics. Instead of
built-in FC polarisation vectors we have used transverse gluon
polarisation vectors which enter the projection in
Eq.~(\ref{GIproj}), and the standard $W^{\pm}$ polarisation vectors
defined in Eq.~(\ref{vectors}), giving us an access to individual
polarisation states of the $W$ bosons. In accordance with the
$k_t$-factorisation technique, the gauge invariance of the resulting
amplitudes for the on-mass-shell initial gluons is ensured by a
projection onto the gluon transverse polarisation vectors
proportional to the transverse gluon momenta $q_{1,2\perp}$
according to Eq.~(\ref{GIproj}).

For the evaluation of the scalar master tree- and four-point
integrals in the gluon-gluon fusion subprocess we have used the {\tt
LoopTools} library \cite{FC}. The result is summed up over all
possible quark flavors in loops and over distinct loop topologies.
We have also checked that the sum of relevant diagrams is explicitly
finite and obeys correct asymptotical properties and energy
dependence. It is worth to mention that a large cancelation between
separate box contributions in the total sum of diagrams takes place,
which is expected from the general Standard Model symmetry
principles
\footnote{We are thankful to Prof. O. Nachtmann for
an enlightening discussion on this matter.}.

As soon as the hard subprocess matrix element (denoted above as
$V_{\lambda_+\lambda_-}$) has been defined as a function of relevant
kinematical variables (four-momenta of incoming/outgoing particles),
the loop integration over $q_{0\perp}$ in Eq.~(\ref{ampl}) was
performed to obtain the diffractive amplitude, which then has been
used to calculate the differential distributions for (un)polarised
$W$ bosons.

As we will demonstrate below, in the Standard Model the total box
contribution is somewhat larger than the triangle one, for the realistic
Higgs boson masses. We, however, keep both the triangle and box
contributions and investigate a possible interference between them,
which, in fact, is quite important, especially at rather large $W^+W^-$-pair
invariant masses, i.e. in the region we are interested in.

\subsection{Exclusive $p p \to p p \gamma \gamma$ process}

\begin{figure}[h!]
\centerline{\epsfig{file=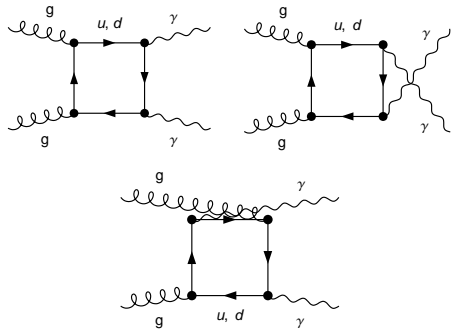,width=6.5cm}} 
\vspace{-2cm}
\caption{Representative diagrams of the hard subprocess $gg\to
\gamma\gamma$, which contribute to the exclusive $\gamma\gamma$ pair
production.} \label{fig:GGhard}
\end{figure}

The same formalism as described above is used to calculate the amplitude
for the $p p \to p p \gamma \gamma$ process.
We write the amplitude of the diffractive 
$p p \to p p \gamma \gamma$ process as
\begin{eqnarray} \label{ampl_gamgam}
{\cal M}_{\lambda_+\lambda_-}(s,t_1,t_2) &\simeq&is\frac{\pi^2}{2}
\int d^2 {\bq}_{0\perp} 
V_{\lambda_+\lambda_-}^{gg \to \gamma \gamma}(q_1,q_2,k_{+},k_{-})
\frac{f_g(q_0,q_1;t_1)f_g(q_0,q_2;t_2)}
{{\bq}_{0\perp}^2\,{\bq}_{1\perp}^2\,{\bq}_{2\perp}^2}\,,
\end{eqnarray}
where now $\lambda_{\pm}=\pm 1$ are the helicity polarisation states 
of the produced photons
and corresponding polarisation vectors are defined
easily in the $pp$ center-of-mass frame 
%
\begin{eqnarray}
\epsilon(k,\pm 1) &=& \frac{1}{\sqrt{2}} \left(0,\, i\sin\phi \mp
\cos\theta\cos\phi,\, -i\cos\phi \mp \cos\theta\sin\phi,\, \pm
\sin\theta\right)\, .
\label{vectors_gamgam}
\end{eqnarray}
The typical contributions to the leading order $gg\to \gamma\gamma$
subprocess are shown in Fig.~\ref{fig:GGhard}. The total number of
topologically different loop diagrams in the Standard Model amounts
to twelve boxes. So the $\gamma\gamma$ does not exhibit resonant
features, and can potentially serve as a probe for New Physics
resonant contributions.

The box contributions to the $gg\to \gamma\gamma$ parton level
subprocess amplitude in Fig.~\ref{fig:GGhard} for on-shell fusing
gluons were calculated analytically by using the Mathematica-based
{\tt FormCalc} (FC) \cite{FC} package. The complete matrix element
was automatically generated by FC tools in terms of one-loop
Passarino-Veltman two-, three- and four-point functions and other
internally-defined functions (e.g. gluon and vector bosons
polarisation vectors) and kinematical variables.

Other details of the calculation are very much the same as those
for the $W^+ W^-$ production. We will not repeat here the details.

\subsection{Gluon $k_{\perp}$-dependent densities in
the forward limit}

In the $k_t$-factorisation approach, the density of gluons in the
proton is described in terms of the off-diagonal unintegrated gluon
distribution functions (UGDFs)
$f_g(q_0,q_{1,2};t_{1,2})=f^{\mathrm{off}}_g(x',x_{1,2},{\bq}_{0\perp}^2,{\bq}_{1/2\perp}^2,\mu_F^2;t_{1,2})$
at the factorization scale $\mu_F \sim M_{WW}\gg |{\bq}_{0\perp}|$.
In the forward scattering (see Eq.~(\ref{forward})) and asymmetric
limit of $x'\ll x_{1,2}$, the off-diagonal UGDF is written as
a skewedness factor $R_g(x')$ multiplied by the diagonal UGDF, which
describes the coupling of gluons with longitudinal momentum
fractions $x_{1,2}$ to the proton (see Refs.~\cite{Kimber:2001sc,MR}
for details). The skewedness parameter $R_g$ is
expected to be roughly constant at LHC energies and gives only a
small contribution to the overall normalization uncertainty.
We take $R_{g} = 1.3$ in practical calculations.
In the kinematics considered here, the unintegrated gluon density can be
written in terms of the conventional integrated gluon distribution
$g(x,{\bq}_{\perp}^2)$ as~\cite{MR}
\begin{eqnarray}\nonumber
f_g(q_0,q_{1,2};t_{1,2})&\simeq&
R_g f_g(x_{1,2},{\bq}_{\perp}^2,\mu_F^2)\exp(bt_{1,2}/2)=\\
&&R_g\frac{\partial}{\partial\ln {\bq}_{\perp}^2}
\Big[x_{1,2} g(x_{1,2},{\bq}_{\perp}^2)\sqrt{T_g({\bq}_{\perp}^2,\mu_F^2)}\Big]
\exp(bt_{1,2}/2)\,, \label{ugdfkmr}
\end{eqnarray}
where the diffractive slope is taken to be $b=4$ GeV$^{-2}$.
$T_g$ is the Sudakov form factor which suppresses real emissions
from active gluons during the evolution,
so that the rapidity gaps are not populated by gluons.
It is given by~\cite{MR}
\begin{eqnarray}
T_g({\bq}_{\perp}^2,\mu_F^2)&=&{\rm exp}
\bigg(-\int_{{\bq}_{\perp}^2}^{\mu_F^2} \frac{d
{\bk}_{\perp}^2}{{\bk}_{\perp}^2}\frac{\alpha_s({\bk}_{\perp}^2)}{2\pi}\int_{0}^{1-\Delta}
\!\bigg[ z P_{gg}(z) + \sum_{q} P_{qg}(z) \bigg]dz \!\bigg)\,,
\label{Sudakov}
\end{eqnarray}
where $\Delta$ in the upper limit is taken to be
\cite{Coughlin:2009tr}
\begin{equation}\label{delta}
\Delta=\frac{|{\bk}_{\perp}|}{|{\bk}_{\perp}|+M_{WW}} \,.
\end{equation}
In our calculations we take $\mu_F^2 = M_{WW}^2$. 
The choice of the scale introduces uncertainties roughly
of about factor two. Since in the present calculations we need
values of $T_g({\bq}_{\perp}^2,\mu_F^2)$ for extremely large scales
$\mu_F^2$ the integration in Eq.~(\ref{Sudakov}) is performed rather
in $\log_{10}(k^2/k_0^2)$, where $k_{0} = 1$ GeV was introduced for convenience.

\section{Four-body phase space in the forward limit}

The diffractive $WW$ CEP amplitude (\ref{ampl}) described above is
used now to calculate the corresponding cross section including
certain limitations of the phase space. The cross section for the
two-boson production can be obtained by integration over the
four-body phase space given by
\begin{eqnarray}
\sigma=\frac{(2 \pi)^4}{2s}\int\overline{ |{\cal M}|^2}\delta^4 (p_1
+ p_2 - p'_1 - p'_2 - k_+ - k_-) \frac{d^3 p'_1}{(2 \pi)^3 2 E'_1}
\frac{d^3 p'_2}{(2 \pi)^3 2 E'_2} \frac{d^3 k_+}{(2 \pi)^3 2 E_+}
\frac{d^3 k_-}{(2 \pi)^3 2 E_-} , \nonumber\\
\label{full_phase_space}
\end{eqnarray}
where $E'_{1,2}$ and $E_{\pm}$ are the energies of the final-state
protons and produced $W^{\pm}$ bosons, respectively, $\overline{
|{\cal M}|^2} = \sum_{\lambda_{+},\lambda_{-}}
{\cal M}_{\lambda_{+}\lambda_{-}} {\cal M}^{*}_{\lambda_{+}\lambda_{-}}$
assuming, as usual, that the helicities of both protons are
unchanged in the considered process. In order to calculate the
total cross section one has to take the eight-dimensional integral
numerically (for details see e.g. Ref.~\cite{LS2010}). However, the
evaluation of the corresponding hard subprocess amplitude
$V_{\lambda_{+}\lambda_{-}}$, its subsequent convolution with the
gluon UPDFs in the diffractive amplitude (\ref{ampl}) and the full
phase space integration (\ref{full_phase_space}) is extremely time
consuming. Clearly the calculation of diffractive mechanism must be
simplified to be feasible. Such a simplification seems possible for
the diffractive process considered here. We start from the choice of
integration variables as in Ref.~\cite{LS2010}. Then
\begin{eqnarray}
d\sigma=\frac{1}{2s}\overline{ |{\cal
M}|^2}\,\frac{1}{2^4}\frac{1}{(2\pi)^8}\frac{1}{E'_1E'_2}\,\frac14\,
dt_1 dt_2 d{\phi_1} d{\phi_2}\,\frac{p_{m\perp}}{4}{\cal
J}^{-1}\,dy_+ dy_- dp_{m\perp} d\phi_m   \, ,
\label{redPS0}
\end{eqnarray}
where $p_{m\perp}=|\bk_{+\perp}-\bk_{-\perp}|$ is the
difference between transverse momenta of $W^+$ and $W^-$,
$\bk_{+\perp}$ and $\bk_{-\perp}$, respectively, and
$\phi_m$ is the corresponding azimuthal angle.
For the sake of simplicity, assuming an exponential slope of
$t_1 / t_2$-dependence of the KMR UGDFs (see Eq.~(\ref{ugdfkmr})), and as a
consequence of the approximately exponential dependence of the cross
section on $t_1$ and $t_2$ (proportional to $\exp(b t_1)$ and
$\exp(b t_2)$), the four-body phase space can be calculated as follows
\begin{eqnarray}
d \sigma \approx && \frac{1}{2s}\overline{
|{\cal M}|^2}\Big|_{t_{1,2}=0}\,\frac{1}{2^4}
\frac{1}{(2\pi)^8}\frac{1}{E'_1E'_2}\,
\frac14\,
\frac{1}{b^2}\,(2\pi)^2\,
\frac{p_{m\perp}}{4}{\cal J}^{-1}\,dy_+
dy_- dp_{m\perp} d\phi_m \,.
\label{redPS}
\end{eqnarray}
Since in this approximation we have assumed no correlations between
outgoing protons (which is expected here and is practically true for
the production of $b \bar b$ \cite{MPS_bbbar} or $g g$
\cite{MPS2011_gg} dijets) there is no dependence of the integrand in
Eq.~(\ref{redPS}) on $\phi_m$, which means that the phase space
integration can be further reduced to three-dimensional one. The
Jacobian ${\cal J}$ in Eq.~(\ref{redPS0}) is given in Ref.~\cite{LS2010}
\begin{eqnarray}
{\cal J}=\Bigg| \frac{p_{1z}'}{\sqrt{m_p^2+{p'}_{1z}^2}} -
\frac{p_{2z}'}{\sqrt{m_p^2+{p'}_{2z}^2}} \Bigg|   \, .
\end{eqnarray}
In actual calculations below we shall use the reduced form of the
four-body phase space Eq.~(\ref{redPS}), and it is checked to give
correct numerical results against the full phase space calculation
for some simple reactions.
Different representations of the phase space depending on a
particular kinematical distributions needed can be found in Ref.~\cite{LS2010}.

\section{$\gamma \gamma \to W^+ W^-$ mechanism}

\begin{figure}[h!]
\centerline{\epsfig{file=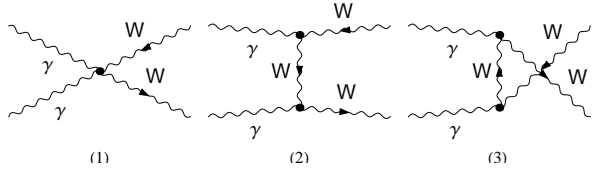,width=8.0cm}}
\caption{The Born diagrams for the $\gamma \gamma \to
W^{\pm}W^{\mp}$ subprocess.}
\label{fig:gamgam_WW}
\end{figure}

In this section, we briefly discuss the $\gamma \gamma \to W^+ W^-$
mechanism, considered already in the literature (see
Refs.~\cite{royon,piotrzkowski}).
The relevant subprocess diagrams are shown in Fig.~\ref{fig:gamgam_WW}.


Let us start from the reminder about the $\gamma \gamma \to W^+ W^-$
coupling within the Standard Model.
The three-boson $WW \gamma$ and four-boson $WW \gamma\gamma$ couplings,
which contribute to the $\gamma \gamma \to W^+ W^-$ process in
the leading order read
\begin{eqnarray}
\lumi_{WW\gamma} & = &
-ie( A_\mu  W^-_\nu \twosidep{\mu} W^{+\nu}
+   W_\mu^- W^+_\nu \twosidep{\mu} A^\nu
+    W^+_\mu  A_\nu \twosidep{\mu} W^{-\nu}) \, ,
\label{eq:anom:lagrww1}\\
\lumi_{WW\gamma\gamma} & = &
-e^2(W^{-}_\mu W^{+\mu}A_\nu A^\nu-W_\mu^-A^\mu W^+_\nu A^\nu) \, ,
\label{eq:anom:lagrww2}
\end{eqnarray}
where the asymmetric derivative has the form
$X\twosidep{\mu}Y=X\p^{\mu}Y-Y\p^{\mu}X$.

Then within the Standard Model, the elementary tree-level cross
section for the $\gamma \gamma \to W^+ W^-$ subprocess can be written in the
compact form in terms of the Mandelstam variables (see e.g. Ref.~\cite{DDS95})
\footnote{This formula does not include the
process with virtual Higgs boson $\gamma \gamma \to H \to W^+ W^-$
\cite{gammagamma_H_WW}. For heavy Higgs boson, this would lead to
clear Higgs boson signal modifying the cross section (typical
resonance $+$ background effect) \cite{DDS95}, however, with the
present limits for Higgs boson mass
\cite{ATLAS_exclusion,CMS_exclusion} only deeply off-shell Higgs
boson contribution could be possible. Also, the diagram with an
intermediate Higgs boson is, of course, of a higher order compared
to the contributions considered here. This automatically means
rather small effect on the measured cross section, in particular, on
the $W^+ W^-$ invariant mass distribution in our case of the
four-body $p p \to p W^+ W^- p$ reaction.}
\begin{equation}
\frac{d\hat{\sigma}}{d \Omega} = \frac{3 \alpha^2 \beta}{2\hat{s}} \left(
1 - \frac{2 \hat{s} (2\hat{s}+3m_W^2)}{3 (m_W^2 - \hat{t}) (m_W^2 -
\hat{u})} + \frac{2 \hat{s}^2(\hat{s}^2+ 3m_W^4)}{3 (m_W^2 -
\hat{t})^2(m_W^2 - \hat{u})^2} \right) \,,
\label{gamgam_WW}
\end{equation}
where $\beta=\sqrt{1-4m_W^2/\hat{s}}$ is the velocity of the $W$
bosons in their center-of-mass frame and the electromagnetic
fine-structure constant $\alpha=e^{2}/(4\pi) \simeq 1/137$ for the
on-shell photon. The total elementary cross section can be obtained
by integration of the differential cross section above.

In the Weizs\"acker-Williams approximation,
the total cross section for the $pp \to pp (\gamma \gamma) \to W^+ W^-$
can be written as in the parton model
\begin{equation}
\sigma = \int d x_1 d x_2 \, f_1^{WW}(x_1) \, f_2^{WW}(x_2) \,
\hat{\sigma}_{\gamma \gamma \to W^+ W^-}(\hat s) \, .
\label{EPA}
\end{equation}
We take the Weizs\"acker-Williams equivalent photon fluxes of protons from Ref.~\cite{DZ}.

To calculate differential distributions the following parton formula
can be conveniently used
\begin{equation}
\frac{d\sigma}{d y_+ d y_- d^2 p_{W\perp}} = \frac{1}{16 \pi^2 {\hat s}^2}
\, x_1 f_1^{WW}(x_1) \, x_2 f_2^{WW}(x_2) \,
\overline{ | {\cal M}_{\gamma \gamma \to W^+ W^-}(\hat s, \hat t, \hat u)
  |^2} \, ,
\label{EPA_differential}
\end{equation}
where momentum fractions of the fusing gluons $x_{1,2}$ are defined in
Eq.~(\ref{xqq}). We shall not discuss here any approach beyond the Standard Model.
A potentially interesting Higgsless scenario of the
$WW$-pair production has previously been discussed e.g. in
Refs.~\cite{royon,piotrzkowski}.

\begin{figure}[!h]
\includegraphics[width=5cm]{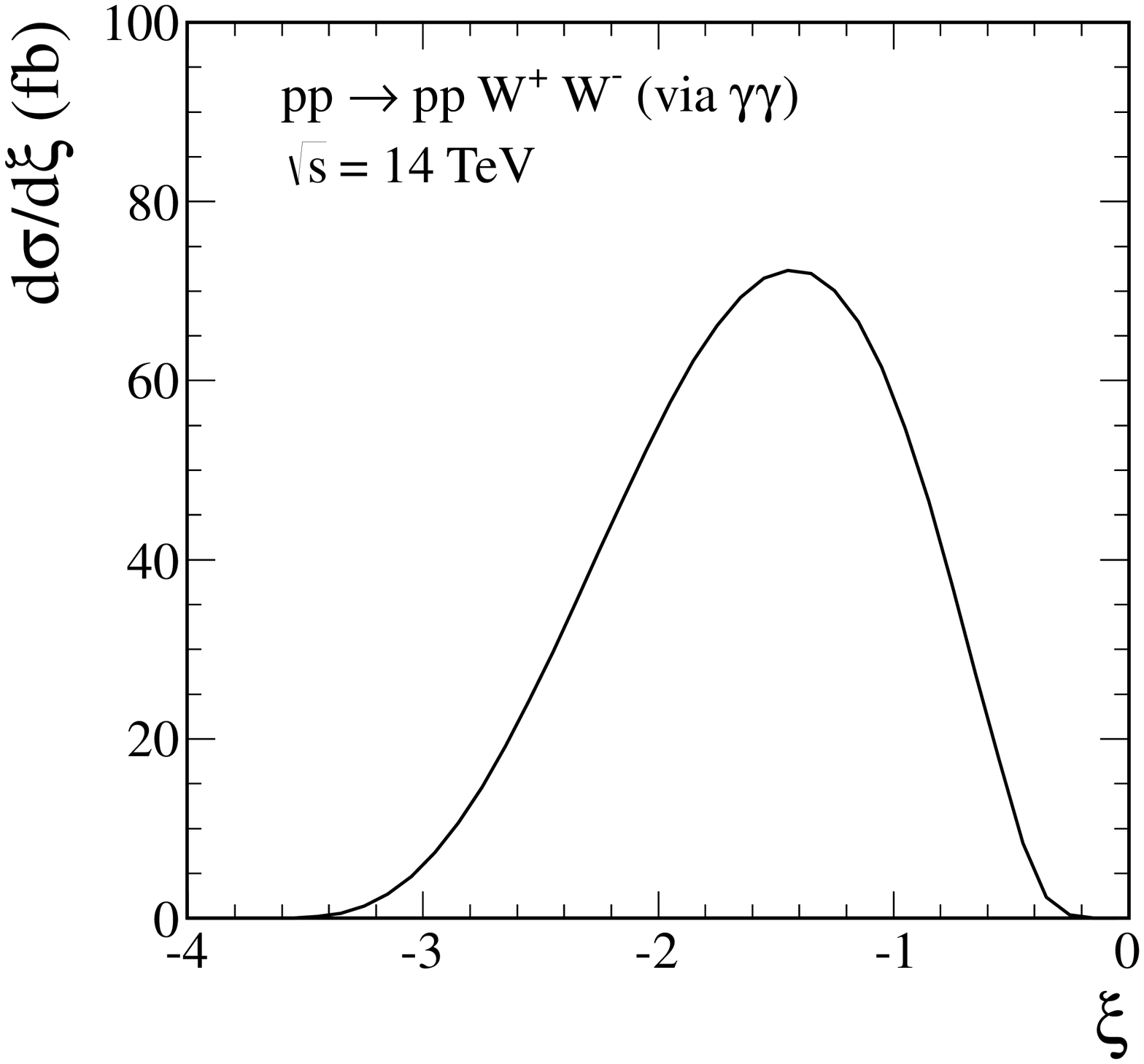}
\includegraphics[width=5cm]{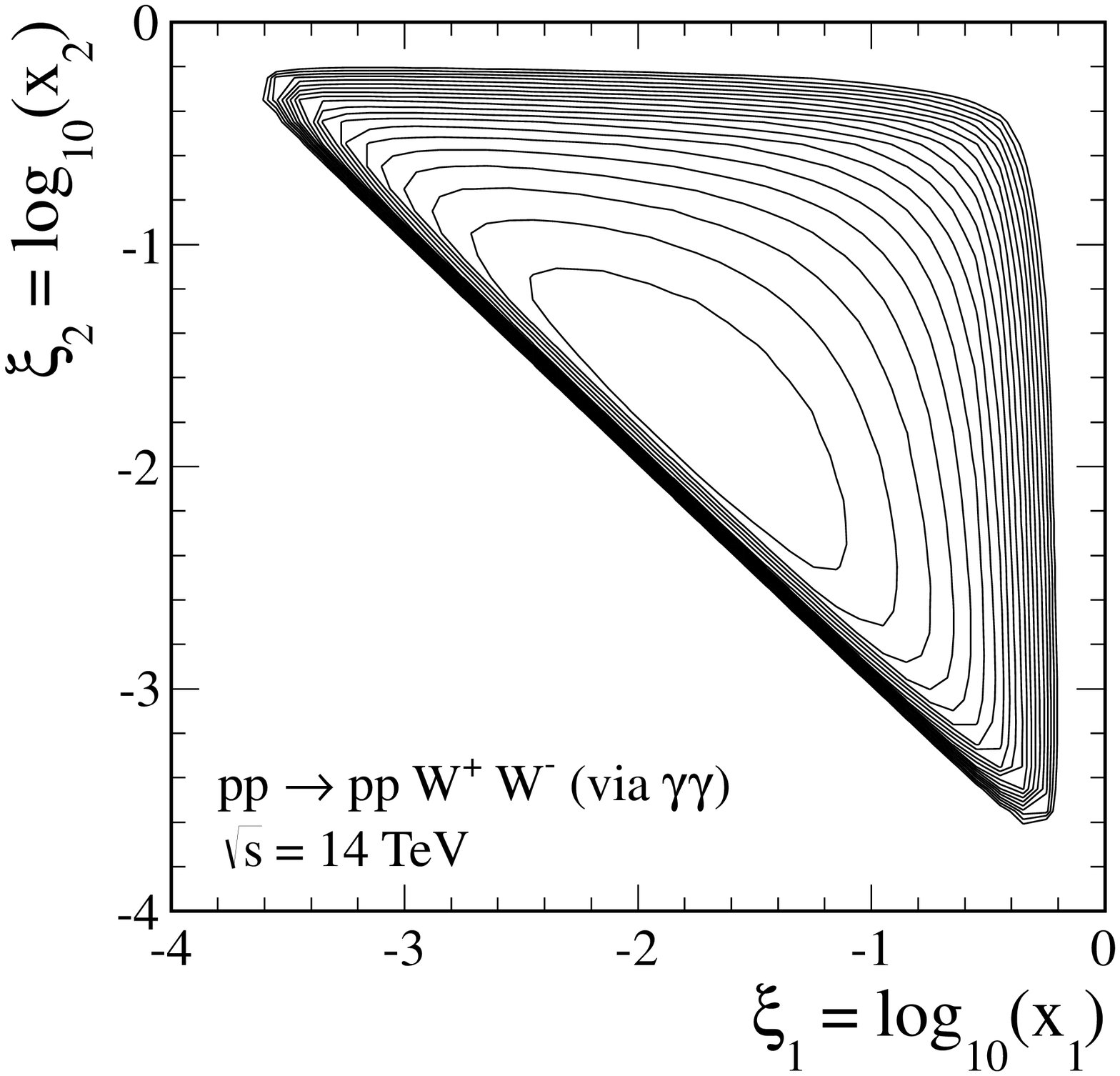}
   \caption{
\small Summary of the $\gamma \gamma \to W^+ W^-$ contribution. The
lines were calculated within EPA approximation as described in the
text with photon fluxes obtained in Ref.~\cite{DZ}. Here, $\xi_{1.2}
= \log_{10}(x_{1,2})$, where $x_{1,2}$ are photon longitudinal
fractions with respect to parent protons.}
 \label{fig:gamma-gamma}
\end{figure}

In Fig.~\ref{fig:gamma-gamma} we show distribution in $\xi_1 =
\log_{10}(x_1)$ and $\xi_2 = \log_{10}(x_2)$ at $\sqrt{s}$ = 14 TeV.
We observe a maximum of the cross section at $\xi_1, \xi_2 \approx
-2$ which means that corresponding longitudinal momentum fractions
carried by photons are typically 10$^{-2}$.

\section{Inclusive production of $W^+W^-$ pairs}

For a test and for a comparison we also consider a gluon-gluon contribution
to the inclusive cross section.
We are not interested in the quark-antiquark component
which is simple and well known. We also omit $pp \to t\bar{t}X \to W^{+}W^{-}b\bar{b}X$
process very important at high energy.
In the lowest order of pQCD the inclusive cross section 
for the gluon-gluon fusion can be written as
\begin{equation}
\frac{d \sigma^{gg}}{d y_+ d y_- d^2 p_{W\perp}} = \frac{1}{16 \pi^2 {\hat s}^2}
x_1 g(x_1, \mu_F^2) x_2 g(x_2,\mu_F^2)
\overline{| {\cal M}_{gg \to W^+ W^-}(\lambda_1, \lambda_2, \lambda_+,
  \lambda_-) |^2} \, .
\label{inclusive_cs}
\end{equation}
The corresponding matrix elements have been discussed in the
literature in detail \cite{inclusive}. The distributions in rapidity
of $W^+$ ($y_+$), rapidity of $W^-$ ($y_-$) and transverse momentum
of one of them $p_{W\perp}$ can be calculated in a straightforward way from
Eq.~(\ref{inclusive_cs}). The distribution in invariant mass can be
then obtained by an appropriate binning.
Our inclusive $d \sigma/ d M_{WW}$ distribution seems consistent with
similar distributions presented in the past in the literature.

\begin{figure}[!h]
\includegraphics[width=7cm]{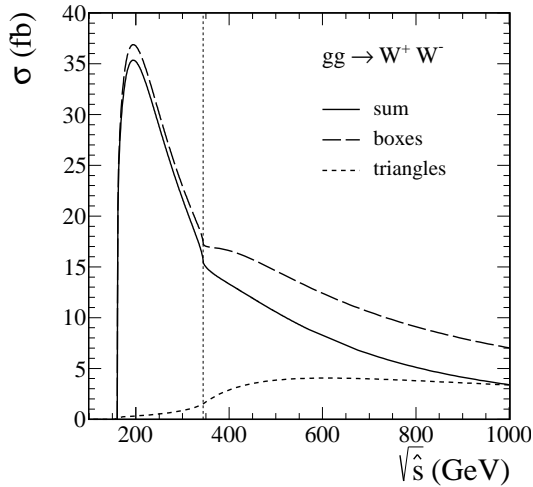}
   \caption{
\small
The integrated elementary cross section for the $gg \to W^{+}W^{-}$ reaction.
The solid line represents the coherent sum of all contributions.
We show separate contributions of boxes (dashed line) and triangles (dotted line).
}
\label{fig:gg_WW}
\end{figure}

The total cross section can be obtained from a simpler formula:
\begin{equation}
\sigma_{pp \to W^{+}W^{-}X}^{gg} = 
\int d x_1 d x_2 \, g(x_1,\mu_{F}^{2}) \, g(x_2,\mu_{F}^{2}) \,
\hat{\sigma}_{gg \to W^+ W^-}(\hat s) \, .
\label{gg_WW}
\end{equation}
Let us concentrate for a while a the elementary $g g \to W^+ W^-$ cross
section shown in Fig.\ref{fig:gg_WW}. 
In this calculation we have assumed $m_{h^{0}}$ = 125 GeV \cite{Higgs}.
We also show a vertical line at the $t \bar t$ threshold.
The figure demonstrates a cancellation
pattern between box and triangle contributions. We will discuss similar
cancellation for the $p p \to p p W^+ W^-$ reaction in the next section.
We wish to notice that
$\hat{\sigma}_{gg \to W^{+}W^{-}} \ll \hat{\sigma}_{\gamma \gamma \to W^{+}W^{-}}
\, {}^{\underrightarrow{\hat{s} \to \infty}} \,\thicksim 10^{2}$ pb.
This shows a potential role of photon-photon induced processes
of $W^+ W^-$ production not discussed
so far in the context of inclusive process.

\begin{figure}[!h]
\includegraphics[width=7cm]{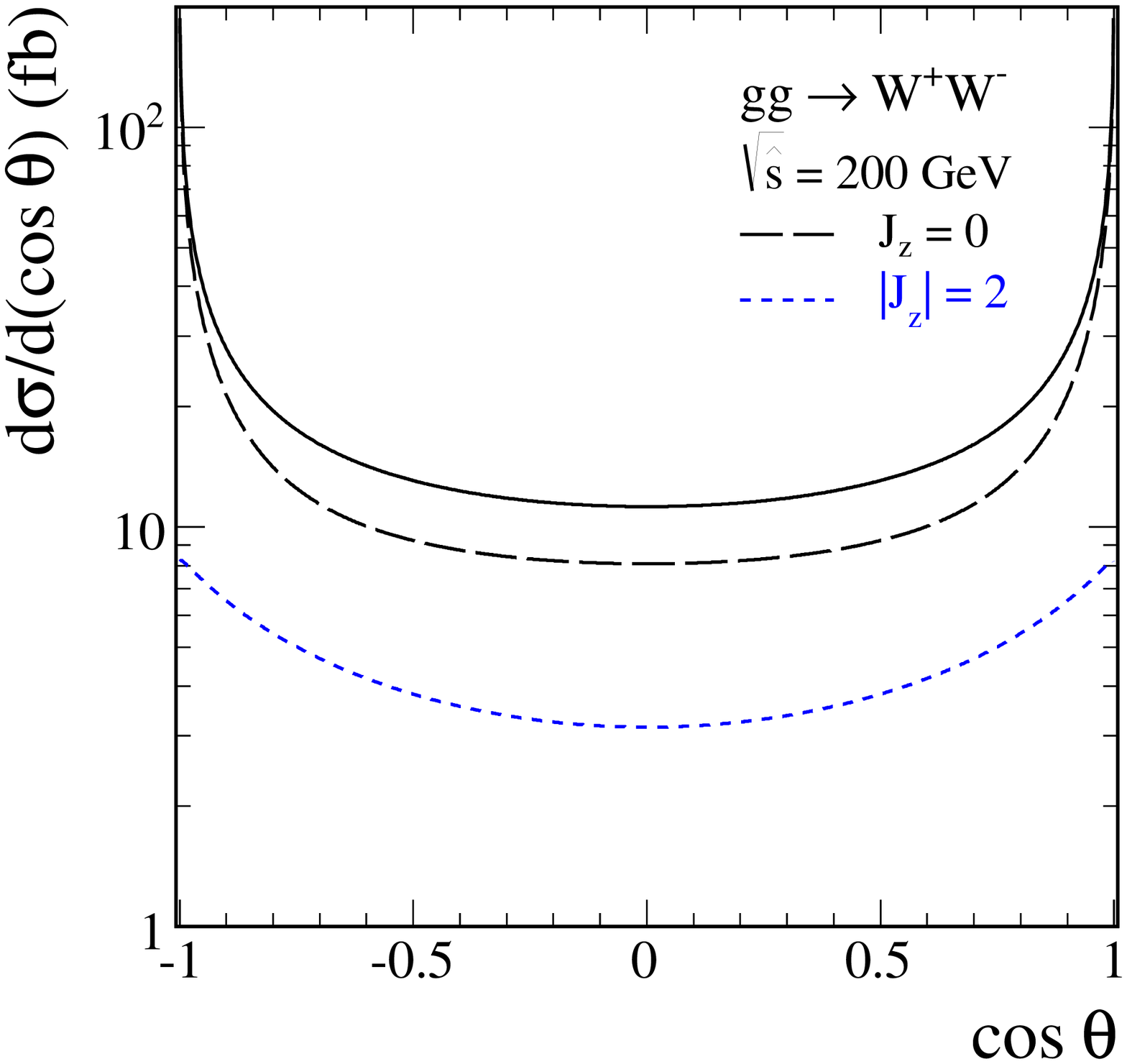}
\includegraphics[width=7cm]{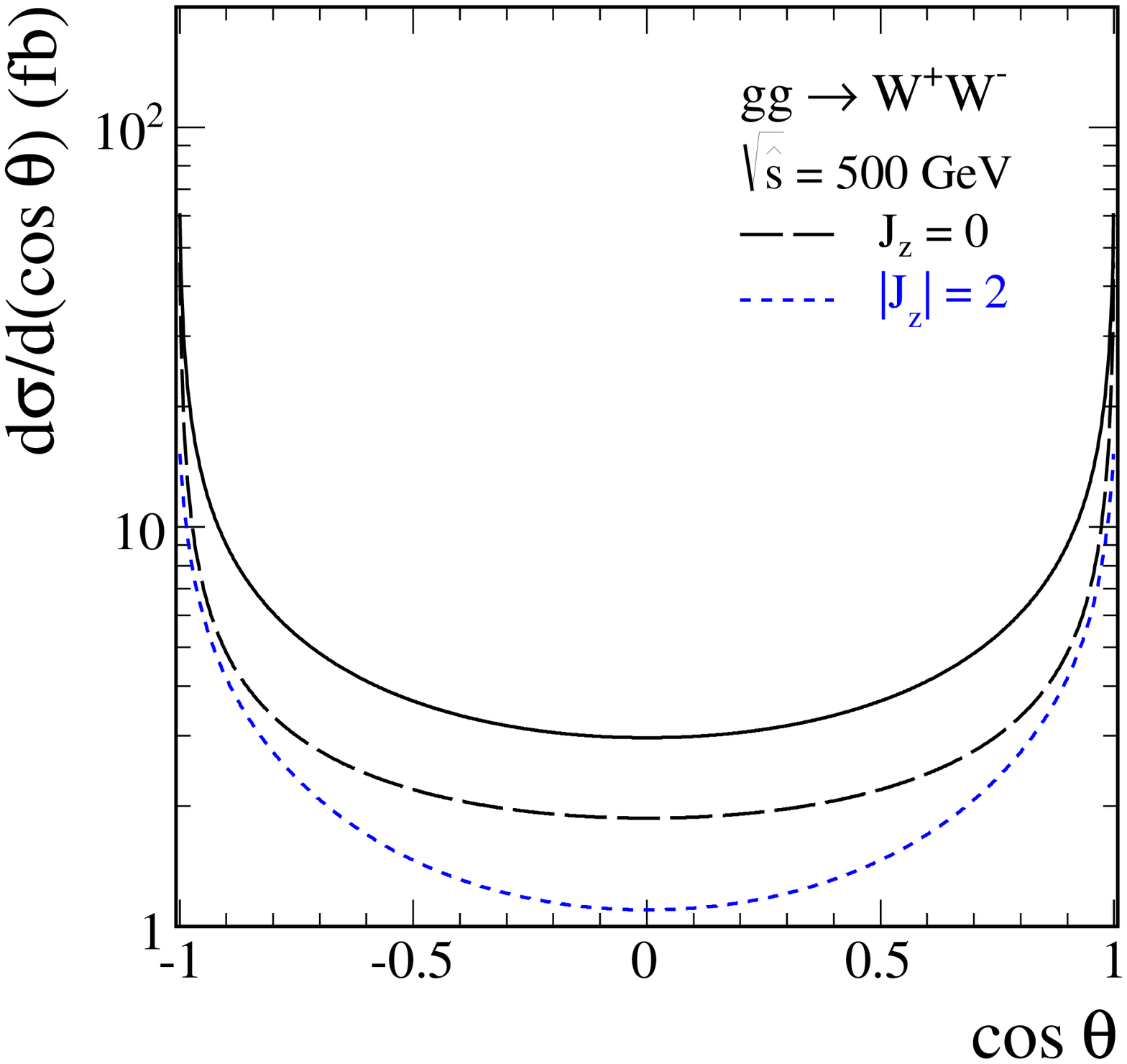}
   \caption{
\small
Centre-of-mass scattering angle dependence of 
the hard subprocess $gg \to W^{+}W^{-}$ cross section
averaged over incoming gluon polarizations.
The solid line represents the coherent sum of all contributions
The $J_{z} = 0$ (dashed line) and the $|J_{z}| = 2$ (dotted line))
contributions are shown separately.
}
\label{fig:ggWW_Jz}
\end{figure}

As discussed before, in the case of exclusive scattering
the $J_z =0$ contribution is the dominant one.
In the case of inclusive process the situation is
slightly different.
In Fig.\ref{fig:ggWW_Jz} we present the $J_{z} = 0$
and $|J_{z}| = 2$ components to angular distributions.
The $J_{z} = 0$ contribution is generally larger than the $|J_{z}| = 2$ one.
As in the exclusive case, at forward scattering ($cos \, \theta = \pm 1$)
we observe the dominance of the $J_z = 0$ contribution.
At $\sqrt{\hat s}$ = 500 GeV it happens very close
to $\cos \, \theta \approx \pm$ 1.

\begin{figure}[!h]
\includegraphics[width=7cm]{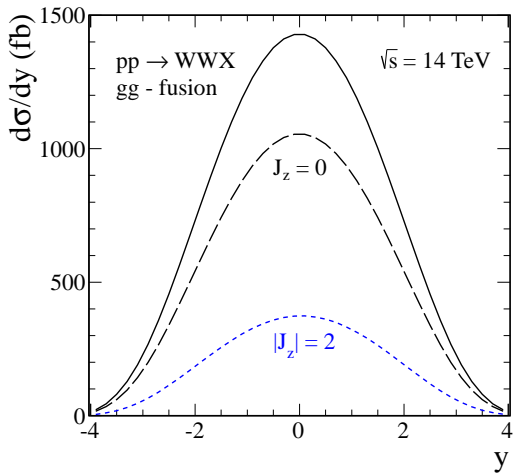}
   \caption{
\small
The $J_z = 0$ (dashed line) and $|J_z| = 2$ (dotted line)
contributions to the inclusive $p p \to W^+ W^- X$
rapidity distribution.
}
\label{fig:pp_WWX}
\end{figure}

For completeness in Fig.\ref{fig:pp_WWX}
we show corresponding contributions to
the rapidity distribution of one of $W$'s
in the $p p \to W^+ W^- X$ process. 
Here the $J_z$ = 0 contribution is larger
in the whole range of rapidities.

\section{Results}

Before we go to the presentation of results for the 
$p p \to p p W^+ W^-$ reaction we wish to show results for
the $p \bar{p} \to p \bar{p} \gamma \gamma$ reaction. 
The latter reaction was
studied experimentally in Ref.\cite{CDF_gamgam}.

\subsection{$p p \to p p \gamma \gamma$}

The $p \bar{p} \to p \bar{p} \gamma \gamma$ process was discussed recently in \cite{HKRS2012}.
No differential distributions have been discussed there.
The CDF Collaboration has measured photons in the interval
$|\eta(\gamma)| <$ 1.0, $E_T >$ 2.5 GeV and with the condition of no
other particles detected in -7.4 $< \eta <$ 7.4. They have obtained
$\sigma_{\gamma \gamma}$ = 2.48 pb with about quarter of relative
uncertainty. We obtain 2.99 pb for the GJR NLO gluon distribution
\cite{GJR}, 2.46 pb for the MSTW08 NLO gluon distribution \cite{MSTW_PDF} 
and 2.1 pb for the CT12 NLO gluon distribution \cite{CT_PDF}. Our results
very well agree with the CDF experimental data. In this calculation we 
have assumed averaged soft gap survival factor $S_g$ = 0.05 and the scale
of the Sudakov form factor was taken as $\mu^2 = M_{\gamma \gamma}^2$. 
Cuts on the gluon transverse momenta $q_{\perp,cut}^{2} = 0.5$ GeV$^{2}$ 
were imposed.

\begin{figure}[!h]
\includegraphics[width=7cm]{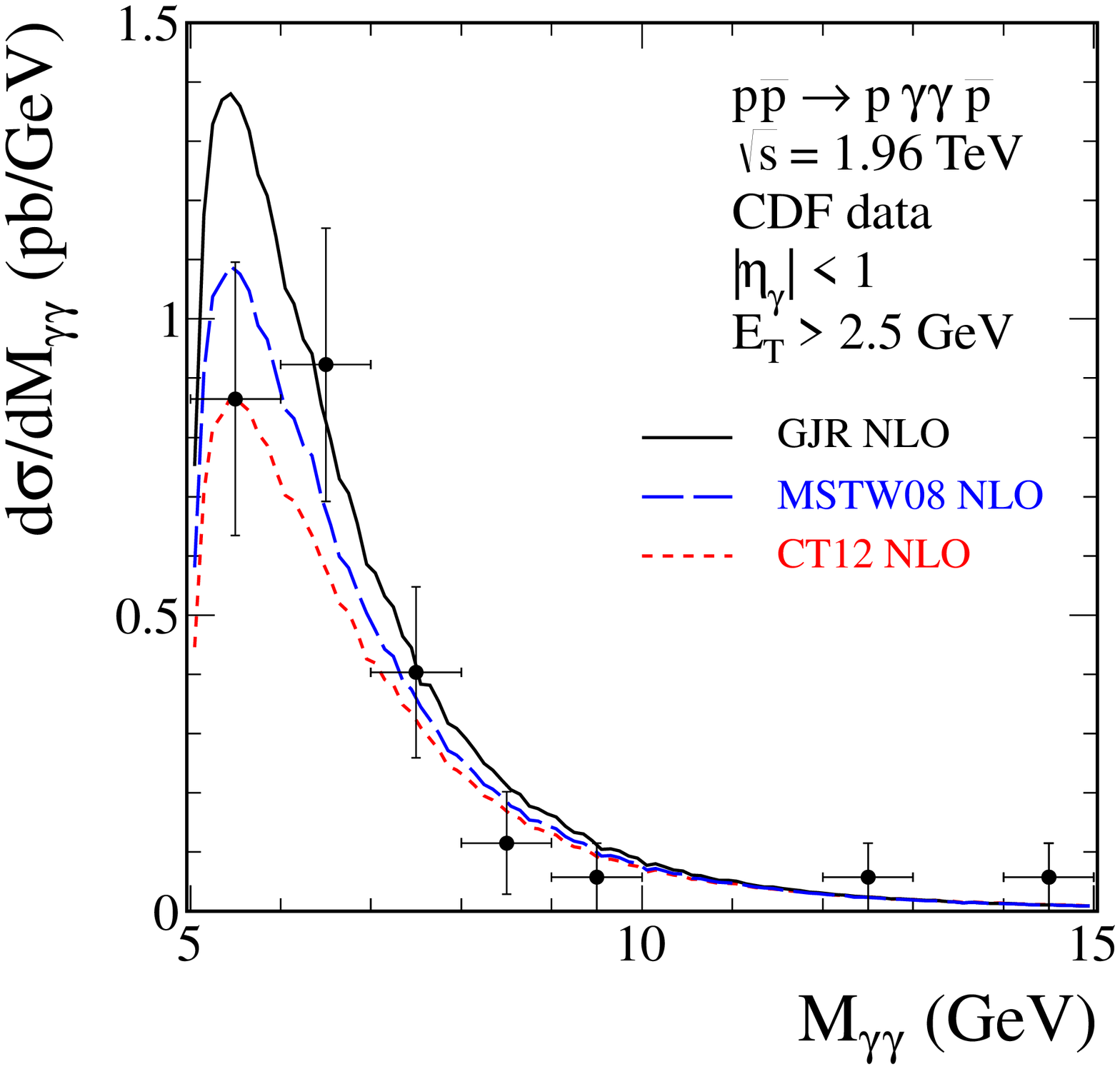}
\includegraphics[width=7cm]{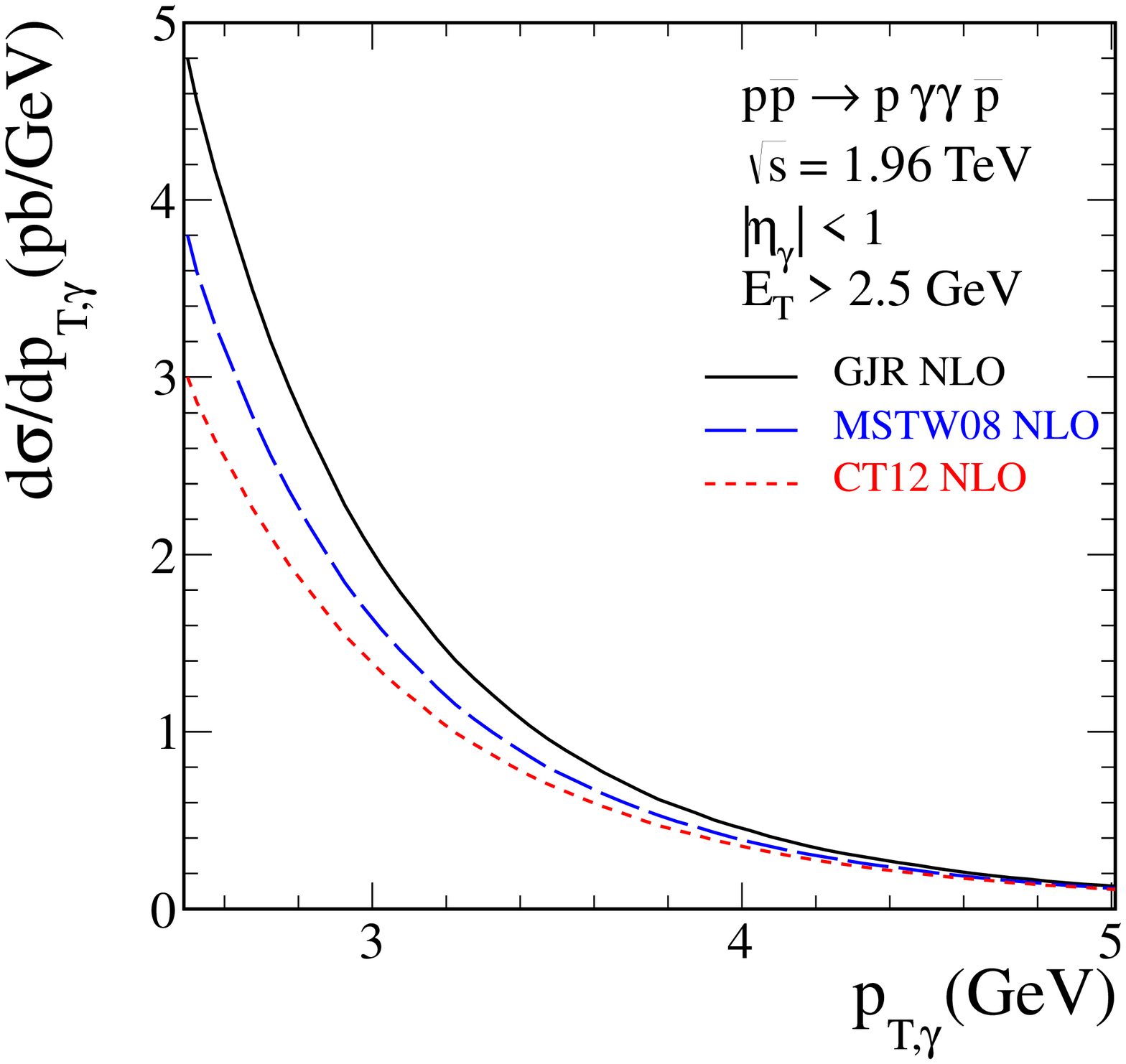}
   \caption{
\small 
Left panel: Photon-photon invariant mass distribution. 
We show results for three different gluon distributions 
specified in the figure. 
The experimental data are taken from Ref.\cite{CDF_gamgam}. 
Right panel: 
}
\label{fig:dsig_dM34_dp1t_gamgam}
\end{figure}

In Fig.\ref{fig:dsig_dM34_dp1t_gamgam} (left panel) we show distribution of photon-photon
invariant mass with experimental CDF cuts. 
We show results for three different gluon
distributions \cite{GJR,MSTW_PDF,CT_PDF}. We obtain very good description
of the CDF experimental data \cite{CDF_gamgam}, both in shape and absolute 
normalization.
In the right panel we show corresponding distribution
in photon transverse momentum.

\begin{figure}[!h]
\includegraphics[width=7cm]{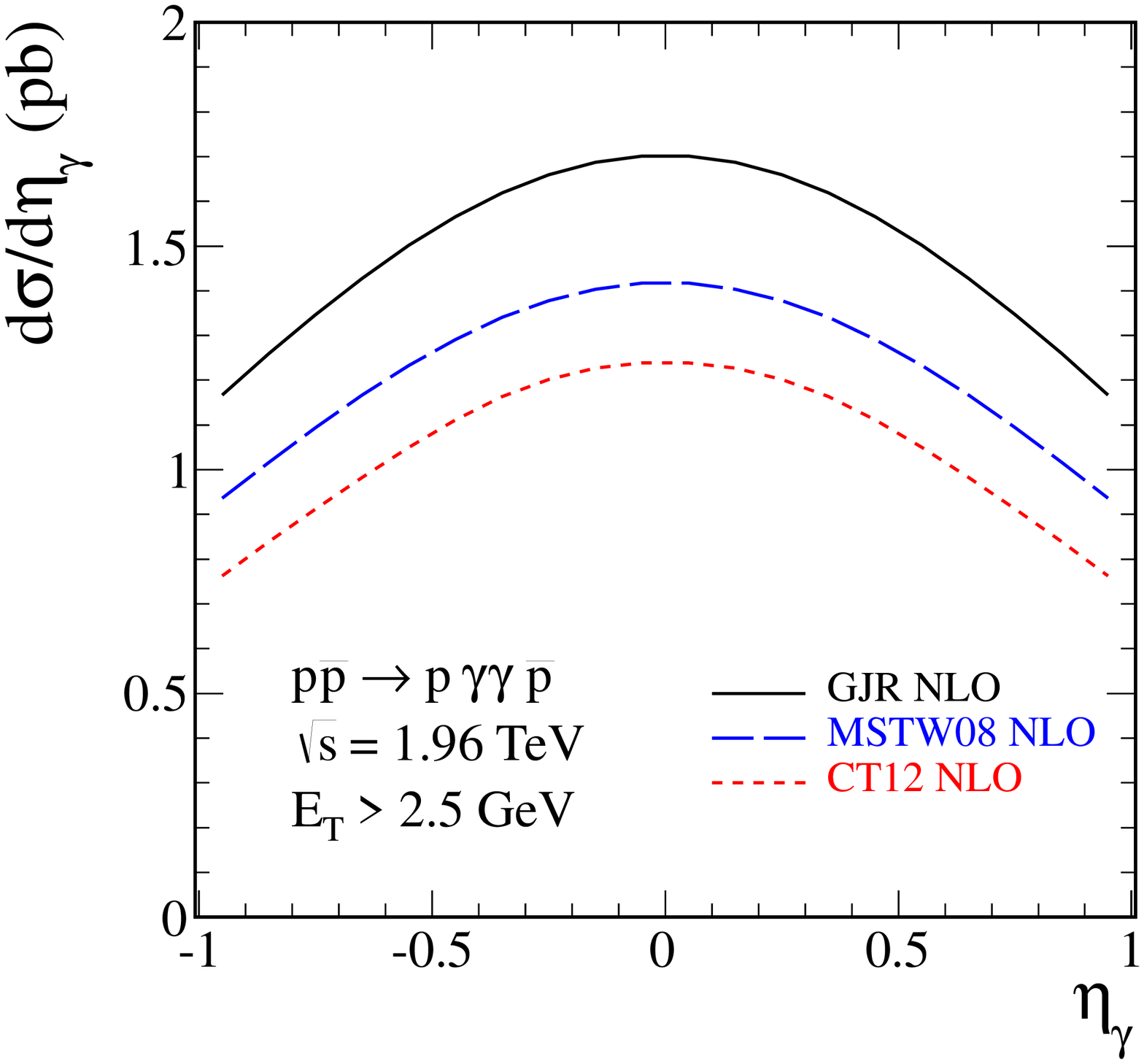}
\includegraphics[width=7cm]{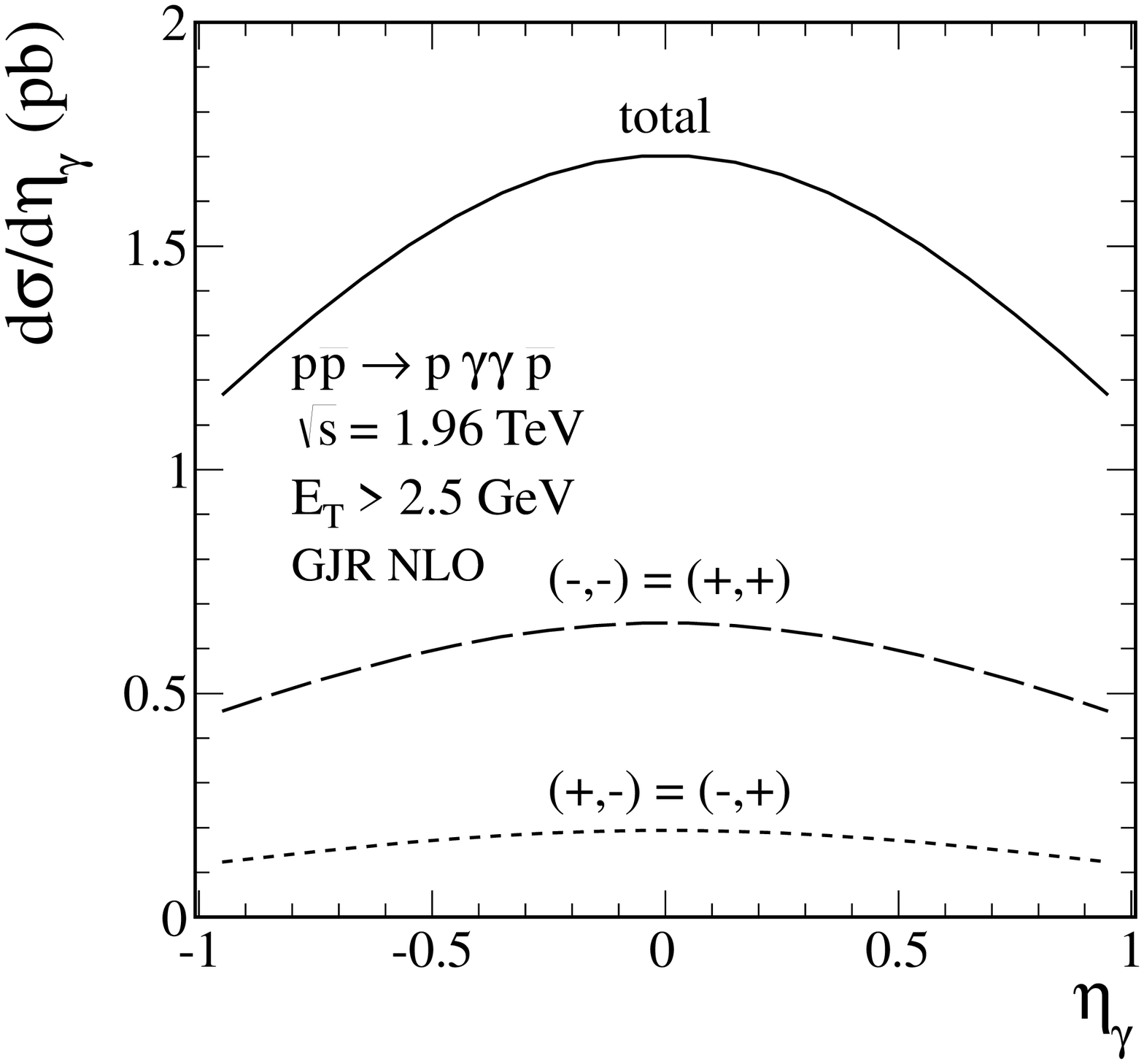}
   \caption{
\small Distribution in photon pseudorapidity for three different
gluon distributions (left panel) and the decomposition into different
$pp$ center-of-mass photon helicity components.
}
\label{fig:dsig_dy_gamgam}
\end{figure}

Finally in Fig.\ref{fig:dsig_dy_gamgam} we show corresponding
distribution in photon pseudorapidity in the left panel, again for three
different gluon distributions. In the right panel we present
decomposition into different $pp$ center-of-mass photon helicity components.

Having shown that the results of the approach used in the present paper
nicely describe the CDF experimental data \cite{CDF_gamgam} we can 
confidentially present our predictions for the $p p \to p p W^+ W^-$ reaction. 

\subsection{$p p \to p p W^+ W^-$}

\begin{figure}[!h]
\includegraphics[width=7cm]{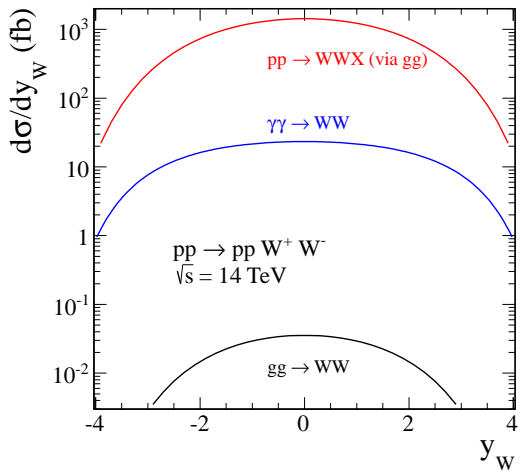}
   \caption{
\small Rapidity distribution of $W$ bosons. The diffractive
contribution is shown by the bottom line while the $\gamma \gamma
\to W^+ W^-$ contribution by the middle line. For comparison, we
also show the cross section for the inclusive ($gg$-fusion only)
production case (upper line).
}
\label{fig:dsig_dy}
\end{figure}

Let us present now our results for the central exclusive $W^+W^-$ pair production.
In Fig.~\ref{fig:dsig_dy} we compare rapidity distribution of $W^+$
(or $W^-$) for the electromagnetic $\gamma \gamma \to W^+ W^-$ and
diffractive $gg \to W^+ W^-$ mechanisms. The two-photon induced
contribution is almost three orders of magnitude larger than the
diffractive contribution, in which all polarization components for
$W^+$ and $W^-$ have been included. For a reference, we show also
inclusive cross section ($g g \to W^+ W^-$ contribution only) which
is roughly two more orders of magnitude bigger than the exclusive 
$\gamma \gamma \to W^+ W^-$ contribution. We see, therefore, that the
exclusive diffractive component is five orders of magnitude smaller
for its inclusive counterpart.
The diffractive contribution was calculated with the GJR NLO
\cite{GJR} collinear gluon distribution, in order to generate the
off-diagonal UGDFs given by Eq.~(\ref{ugdfkmr}). This collinear PDF allows us to
use quite small values of gluon transverse momenta ($q_{\perp,cut}^{2}$ = 0.5 GeV$^2$).

A much smaller diffractive contribution compared to the two-photon
one requires a special comment as it is rather exceptional. 
For example, it is completely opposite than for $p p \to p p H$ \cite{MPS_bbbar}, 
$p p \to p p M$ (e.g. light/heavy quarkonia production \cite{chic,LKRS10}) or
$p p \to p p Q\bar Q$ \cite{MPS_bbbar,MPS_ccbar} CEP processes.
The standard relative suppression,
present also in the latter cases, is due to soft gap survival probability factor
($S_g \sim$ 0.03 for diffractive contribution versus $S_g
\sim$ 1 for two-photon contribution), and due to a suppression by
the Sudakov form factor calculated at very large scales, here at
$\mu_{hard} = M_{WW}$. The main difference compared to other cases is that
in the diffractive case the leading contribution comes from loop
diagrams while in the two-photon case already from tree level
diagrams.

\begin{figure}[!h]
\includegraphics[width=7cm]{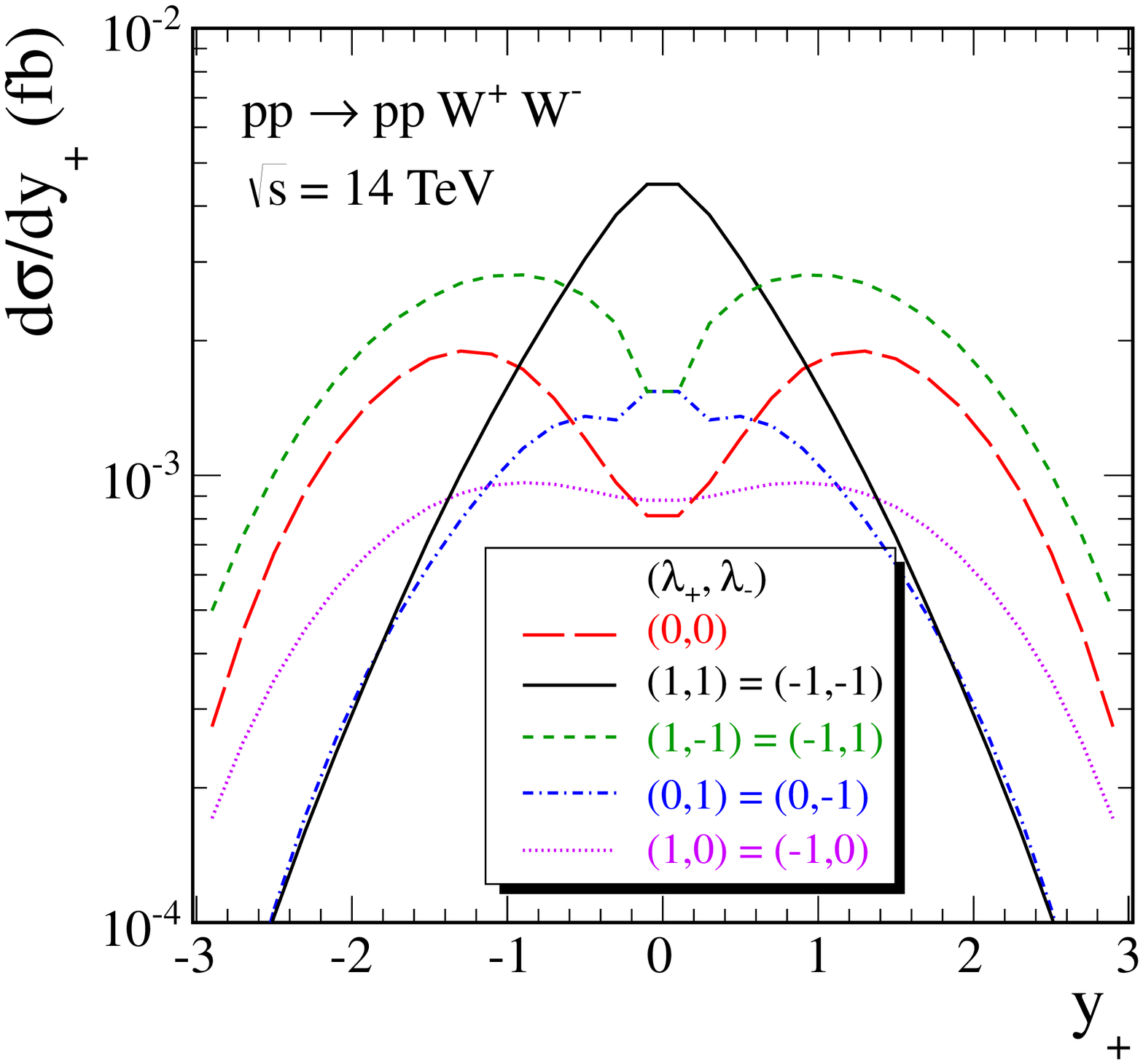}
\includegraphics[width=7cm]{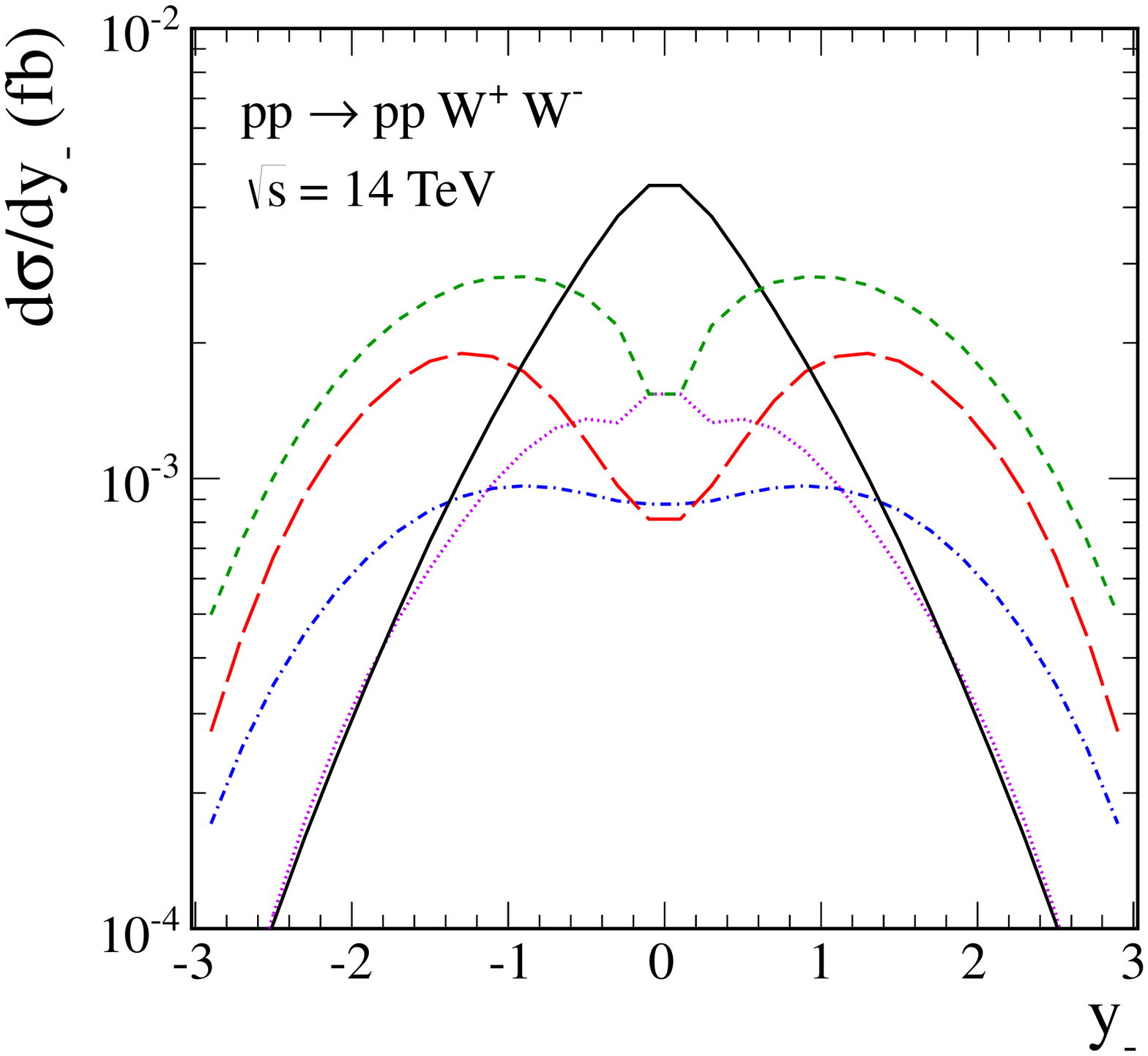}
   \caption{
\small  Rapidity distribution of separate polarisation components to
the diffractive $W$ bosons production. The individual contributions
are marked in the figure.
}
\label{fig:dsig_dy_deco}
\end{figure}

In Fig.~\ref{fig:dsig_dy_deco} we present, in addition, individual
polarization components for the diffractive mechanism, along with
the unpolarized cross section. The calculation of the helicity
contributions is performed in the $pp$ center-of-mass frame (in
which all the experimental studies of the exclusive production
processes are usually performed). 
As can be seen from the figure,
the contribution of $(\lambda_{+},\lambda_{-}) = (\pm 1, \mp 1)$ is
bigger than other contributions and 
the contribution of $(\lambda_{+},\lambda_{-}) = (\pm 1, \pm 1)$
concentrated mostly at midrapidities. Since we use
$p p$ center of mass helicities there is on simple relation
to the often used in a qualitative discussion $J_z$ = 0 dominance rule.
Discussion of the $J_z$ = 0 rule would require complicated
transformations between different reference frames and
going beyond approximations made here. 
This clearly goes beyond the scope of this paper.
In particular, as it is seen from
Fig.~\ref{fig:dsig_dy_deco} the helicity contributions obey
the following relation
\begin{eqnarray}
\frac{d\sigma_{\lambda\lambda'}(y_+)}{dy_+}=\frac{d\sigma_{\lambda'\lambda}(y_-)}{dy_-} \,,
\end{eqnarray}
where $y_{\pm}$ are rapidities of $W^{\pm}$ bosons, respectively.
The unpolarized cross section does not show up any peculiarities in
$y$-dependence and is symmetric with respect to $y=0$ for both $W^+$
and $W^-$ bosons.

\begin{figure}[!h]
  \includegraphics[width=7cm]{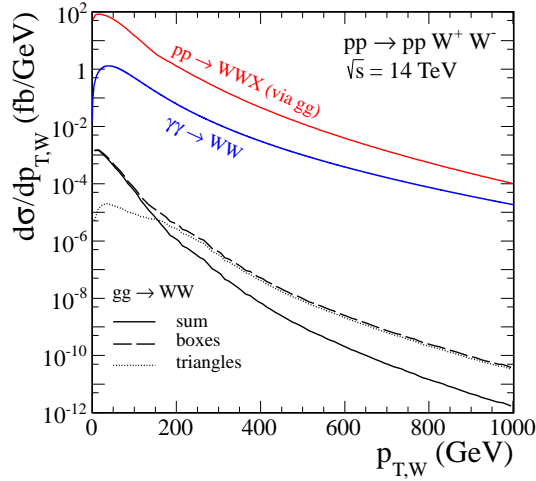}
   \caption{
\small Distribution in transverse momentum of one of the $W$ bosons.
The diffractive contribution is shown by the bottom solid line while
the $\gamma \gamma \to W^+ W^-$ contribution by the middle solid
line. The top solid line corresponds to the inclusive
two-gluon initiated $pp \to W^+W^-X$ component. Separate contributions
of boxes (dashed) and triangles (dotted) are shown in addition for 
illustrating the cancellation effect.
}
 \label{fig:dsig_dpt}
\end{figure}

In Fig.~\ref{fig:dsig_dpt} we show distribution in $W^+$ ($W^-$) transverse momentum.
The distribution for exclusive diffractive production is much
steeper than that for the electromagnetic contribution.
A side remark is in order here.
The diffractive contribution peaks at $p_{t,W} \sim$ 25 GeV.
This is somewhat smaller than for the $\gamma \gamma \to W^+ W^-$ mechanism
where the maximum is at $p_{t,W} \sim$ 40 GeV.
The exclusive cross section for photon-photon
contribution is at large transverse momenta $\sim$ 1 TeV smaller
only by one order of magnitude than the inclusive $gg \to W^{+}W^{-}$ component.
The situation could be even more
favorable if New Physics would be at the game \cite{royon}.

\begin{figure}[!h]
 \includegraphics[width=7cm]{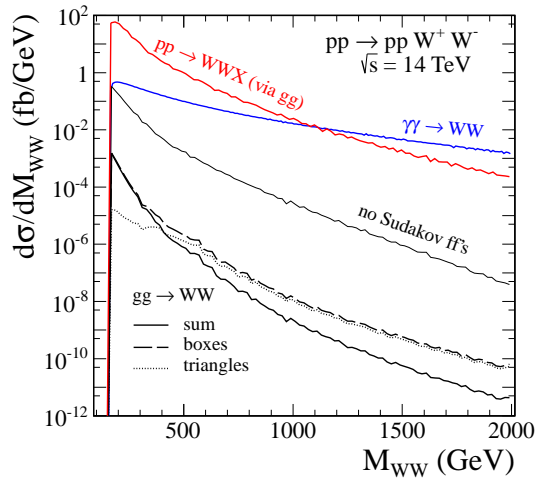}
   \caption{
\small Distribution in $W^+W^-$ invariant mass. We show both the QCD
diffractive contribution and the electromagnetic $\gamma \gamma \to
W^+ W^-$ contribution. The result when the Sudakov form factor is
put to one is shown for illustration of its role. The most upper
curve is for the inclusive gluon-initiated $pp \to W^+ W^-X$ component.}
\label{fig:dsig_dMWW}
\end{figure}

Fig.~\ref{fig:dsig_dMWW} shows distribution in the $W^+ W^-$ invariant
mass which is particularly important for the New Physics searches at
the LHC \cite{royon}. The distribution for the diffractive component
drops quickly with the $M_{WW}$ invariant mass. For reference and
illustration, we show also distribution when the Sudakov form
factors in Eq.~(\ref{ugdfkmr}) is set to one. As can be seen from
the figure, the Sudakov form factor lowers the cross section by a
large factor. The damping is $M_{WW}$-dependent as can be seen by
comparison of the two curves. The larger $M_{WW}$ the larger the
damping. We show the full result (boxes + triangles) and the result
with boxes only which would be complete if the Higgs boson
does not exist. At high invariant masses, the interference of boxes
and triangles decreases the cross section. The distribution for the
photon-photon component drops very slowly with $M_{WW}$ and at
$M_{WW} >$ 1 TeV the corresponding cross section is even bigger than
the $gg \to W^+ W^-$ component to inclusive production of $W^+ W^-$
pairs.

\begin{figure}[!h]
 \includegraphics[width=6cm]{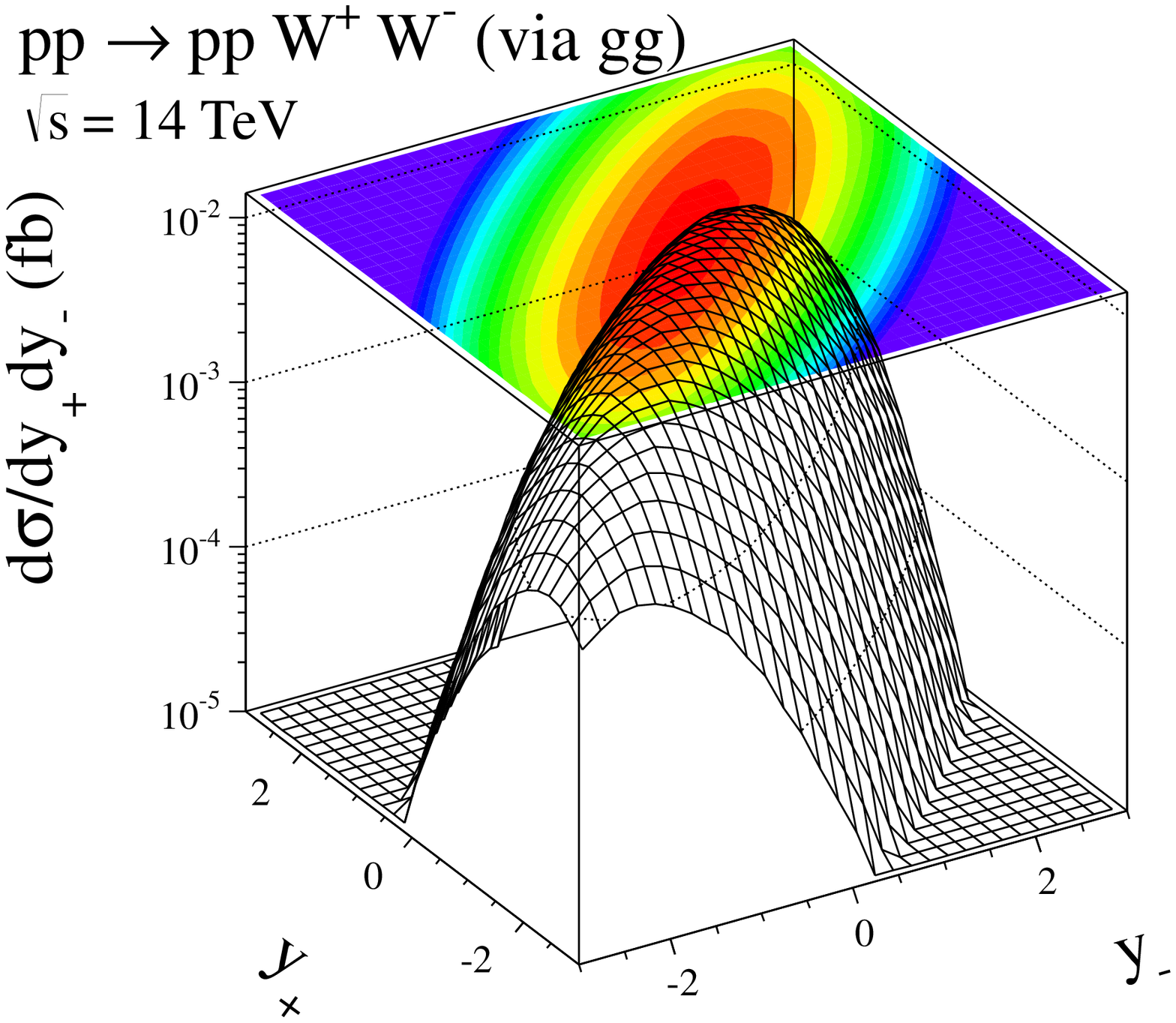}
 \includegraphics[width=6cm]{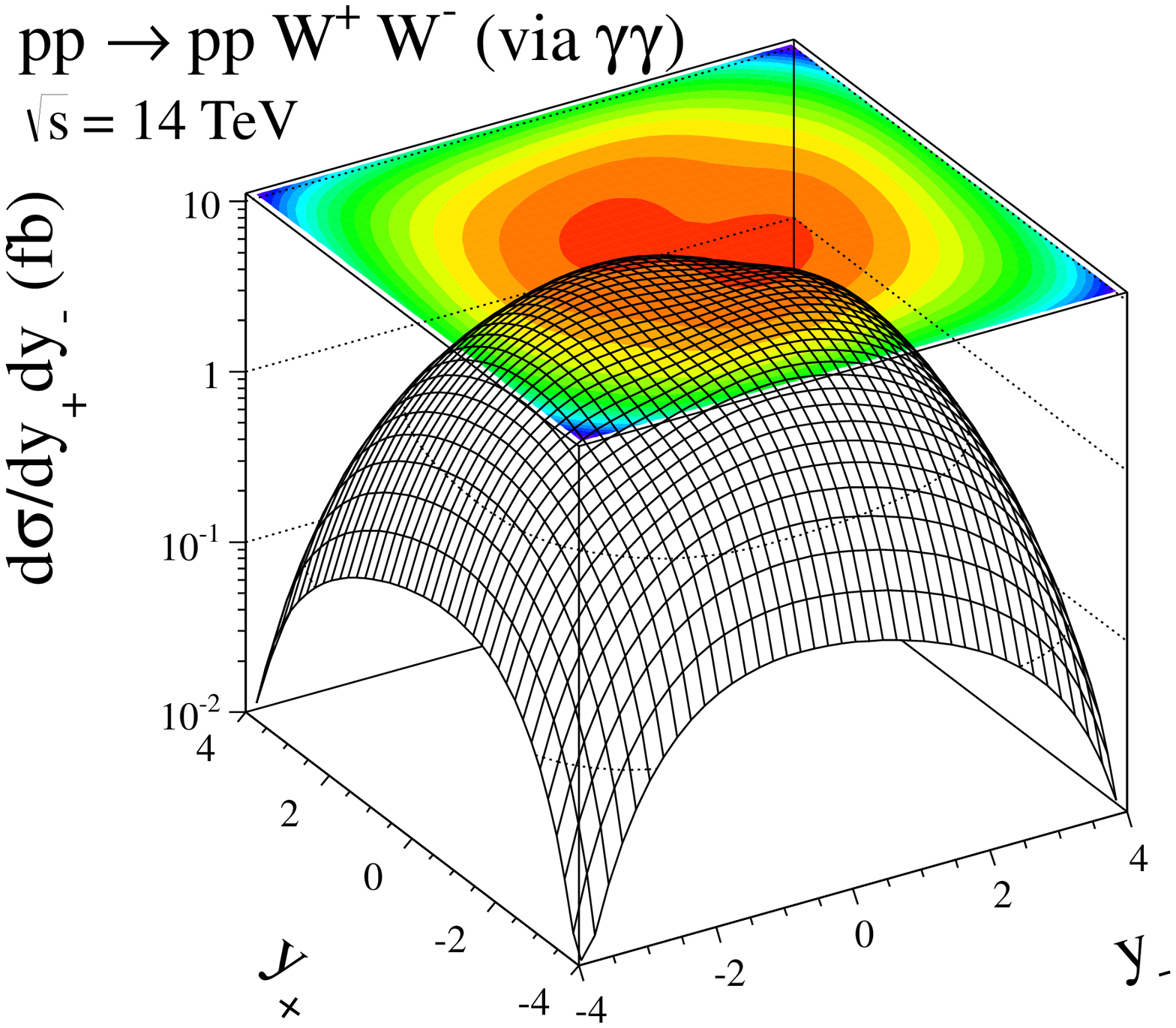}
   \caption{
\small
Two-dimensional distribution in rapidity of $W^+$ and $W^-$ bosons
for the diffractive mechanism (left panel) and two-photon mechanism
(right panel).
}
 \label{fig:dsig_dy1dy2}
\end{figure}

Finally, in Fig.~\ref{fig:dsig_dy1dy2} we show for completeness the
two-dimensional distributions in rapidities of $W^+$/$W^-$ bosons in
both electromagnetic and QCD mechanisms.
We see a typical correlation pattern characteristic for 2 $\to$ 2 subprocesses.
This distribution does not show any specific behavior which could be
used to differentiate the diffractive and the two-photon contributions.

\section{Conclusions}

We have calculated the QCD diffractive contribution to the exclusive
$p p \to p W^+ W^- p$ process for the first time in the literature
with the full one-loop $gg\to W^+W^-$ matrix element. Two mechanisms
have been considered. First mechanism is a virtual (highly off-shell)
Higgs boson production and its subsequent transformation into real
$W^+ W^-$ pair. Second mechanism relies on the formation of
intermediate quark boxes, very much similar to ones in the exclusive
two photon production mechanism.

We have calculated corresponding amplitudes using computer
program package {\tt FormCalc}. We have made a first estimate of
the cross section using amplitudes in the forward limit ``corrected''
off-forward via a simple exponential (slope dependent) extrapolation.

In order to gain confidence to our calculations and the formalism used we
consider also the $p \bar{p} \to p \bar{p} \gamma \gamma$ process
which was measured recently by the CDF Collaboration. 
Here the formalism of calculating
quark box diagrams is essentially the same 
as for the exclusive production of 
$W^+ W^-$ pairs. We have obtained very nice agreement with experimental
diphoton invariant mass distribution.

Having verified the formalism for diphoton production we have performed 
similar calculation for $W^+ W^-$ production.
Differential distributions in the $W^{\pm}$ transverse momentum,
rapidity and $W^+ W^-$ pair invariant mass have been calculated and
compared with corresponding distributions for discussed in the
literature $\gamma \gamma \to W^+ W^-$ mechanism.
The contribution of triangles with the intermediate Higgs boson
turned out to be smaller
than the contribution of boxes taking into account recent
very stringent limitations on Higgs boson mass from Tevatron and LHC data.
We have found that, in contrast to exclusive production of
Higgs boson or dijets, the two-photon fusion dominates over the
diffractive mechanism for small four-momentum transfers squared in
the proton lines ($t_1, t_2$) as well as in a broad range of $W^+
W^-$-pair invariant masses, in particular, for large $M_{WW}$.
Estimated theoretical uncertainties cannot disfavor this statement.
The large $M_{WW}$ region is damped in the diffractive model via
scale dependence of the Sudakov form factor.

One could focus on the diffractive contribution by imposing lower
cuts on $t_1$ and/or $t_2$ using very forward detectors
on both sides of the interaction point at distances of
220 m and 420 m as planned for future studies at ATLAS and CMS. The
corresponding cross section is, however, expected to be extremely low.

Compared to the previous studies in the effective field theory
approach, in this work we have included the complete one-loop
(leading order) $gg\to W^+W^-$ matrix element, and have shown that
extra box diagrams, even though they are larger than the resonant
(s-channel Higgs) diagrams, constitute a negligibly small background for
a precision study of anomalous couplings.

The unique situation of the dominance of the $\gamma \gamma \to W^+
W^-$ contribution over the diffractive one opens a possibility of
independent tests of the Standard Model as far as the triple-boson
$\gamma W W$ and quartic-boson $\gamma \gamma W W$ coupling
is considered. It allows also for stringent tests of some Higgsless
models as discussed already in the literature (see e.g.
Ref.~\cite{royon}).

\section{Acknowledgments}

Useful discussions with Rikard Enberg, Gunnar Ingelman,
Valery Khoze, Otto Nachtmann, Christophe Royon, Torbj\"{o}rn Sj\"{o}strand
and Marek Tasevsky are gratefully acknowledged. This study was partially
supported by the MNiSW grants No. DEC-2011/01/N/ST2/04116 and
DEC-2011/01/B/ST2/04535. Authors are grateful to the European Center
for Theoretical Studies in Nuclear Physics and Related Areas
(ECT$^*$, Trento, Italy) for warm hospitality
during their stay when this work was completed.
Piotr Lebiedowicz is thankful to the THEP Group at
Lund University (LU), Sweden, for hospitality during his
collaboration visit at LU.


\end{document}